\documentclass[aip,jcp,reprint,amsmath,amssymb,nobalancelastpage]{revtex4-2}
\usepackage{graphicx}
\usepackage[suffix=]{epstopdf}
\usepackage{natmove}
\usepackage{multirow}
\newcommand{\vv}[1]{\mathbf{#1}}
\renewcommand{\d}[1]{\ensuremath{\operatorname{d}\!{#1}}}

\begin{document}
\title{Dynamic density functional theory for drying colloidal suspensions: Comparison of hard-sphere free-energy functionals}

\author{Mayukh Kundu}
\affiliation{Department of Chemical Engineering, Auburn University, Auburn, AL 36849, USA}

\author{Michael P. Howard}
\email{mphoward@auburn.edu}
\affiliation{Department of Chemical Engineering, Auburn University, Auburn, AL 36849, USA}

\begin{abstract}
Dynamic density functional theory (DDFT) is a promising approach for predicting the structural evolution of a drying suspension containing one or more types of colloidal particles. The assumed free-energy functional is a key component of DDFT that dictates the thermodynamics of the model and, in turn, the density flux due to a concentration gradient. In this work, we compare several commonly used free-energy functionals for drying hard-sphere suspensions including local-density approximations based on the ideal-gas, virial, and Boubl\'{i}k--Mansoori--Carnahan--Starling--Leland (BMCSL) equations of state as well as a weighted-density approximation based on fundamental measure theory (FMT). To determine the accuracy of each functional, we model one- and two-component hard-sphere suspensions in a drying film with varied initial heights and compositions, and we compare the DDFT-predicted volume-fraction profiles to particle-based Brownian dynamics (BD) simulations. FMT accurately predicts the structure of the one-component suspensions even at high concentrations and when significant density gradients develop, but the virial and BMCSL equations of state provide reasonable approximations for smaller concentrations at a reduced computational cost. In the two-component suspensions, FMT and BMCSL are similar to each other but modestly overpredict the extent of stratification by size compared to BD simulations. This work provides helpful guidance for selecting thermodynamic models for soft materials in nonequilibrium processes such as solvent drying, solvent freezing, and sedimentation.
\end{abstract}

\maketitle

\section{Introduction}
Coatings and films of colloidal particles enable a variety of important technologies, ranging from common consumer products such as latex paints \cite{Keddie:1997,Keddie:2010,Routh:2013} to advanced materials with abrasion \cite{tinkler:jcs:2020} or bacteria \cite{dong:acs:2020} resistance. A common strategy to form such coatings is to suspend colloidal particles in a volatile solvent, cast a film of the suspension onto a substrate, and dry the solvent to concentrate the particles into a solid. The internal distribution of suspended particles can vary as the film dries \cite{Routh:1998,Routh:2004}. For example, particles will concentrate near the solvent--air interface if they collect at the interface faster than they diffuse away, creating a time-dependent concentration gradient. The conditions that produce gradients can be estimated using the film P\'{e}clet number ${\rm Pe} = v H_0/D$ \cite{Routh:2004}, where $v$ is the typical speed at which the solvent--air interface recedes, $H_0$ is the initial film height, and $D$ is a typical particle diffusion coefficient. The particles remain mostly uniformly distributed (close to equilibrium) when ${\rm Pe} \ll 1$, while gradients develop (out of equilibrium) when ${\rm Pe} \gg 1$.

Concentration gradients during drying have recently been recognized to be particularly important for multicomponent coatings \cite{Schulz:2018}. Compositionally distinct layers were found to spontaneously assemble (``stratify'') during fast drying (${\rm Pe} \gg 1$), producing small-on-top \cite{Fortini:2016,MartinFabiani:2016fj,Makepeace:2017}, big-on-top \cite{Trueman:2012ve}, and ``sandwich'' layers \cite{liu:jcis:2019} of two differently-sized particles. Such structures would typically be inaccessible at equilibrium; however, they can form by nonequilibrium mechanisms and become kinetically arrested at solid packing. There is considerable interest to engineer colloidal suspensions and drying processes in order to efficiently fabricate structured coatings, e.g., with a functional component enriched at the surface \cite{tinkler:jcs:2020,dong:acs:2020}.

\begin{figure}
    \includegraphics{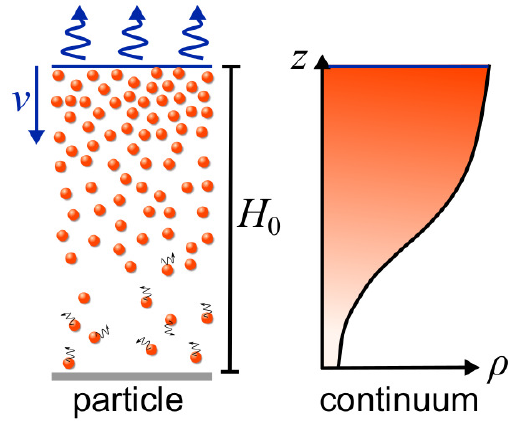}
    \caption{Schematic of particle-based models (left) and continuum models (right) for drying films.}
    \label{fig:schematic}
\end{figure}

Exploring the many chemical and physical parameters that affect film formation is a time-consuming experimental task \cite{Routh:2013,Schulz:2018}, so accurate computational models are highly desired. Most models in prior work on drying colloidal suspensions can be divided into two categories: particle-based simulations \cite{cheng:jcp:2013,howard:jcp:2018,wang:sm:2017,Fortini:2016,Fortini:2017,howard:lng:2017,howard:lng:2017b,tang:lng:2019,tang:jcp:2019,liu:acsnano:2019,chun:sm:2019,Park:2022,song:fpe:2021} and continuum models \cite{Routh:2004,Fortini:2016,howard:lng:2017,howard:lng:2017b,zhou:prl:2017,Trueman:2012er,sear:pre:2017,sear:jcp:2018,reeszimmerman:jfm:2021} (Fig.~\ref{fig:schematic}). Particle-based simulations use techniques such as molecular dynamics or Langevin dynamics to model the trajectories of individual colloidal particles \cite{Allen:2017}. Having particle-level resolution of the evolving film structure is powerful but computationally demanding, making it challenging to capture experimentally relevant particle sizes, film heights, and drying times. Continuum models, largely based on diffusion equations \cite{deGroot:1963,Brady:2011bh}, are attractive alternatives that smooth over particle-level detail and capture variations in the local average particle density. As such, continuum models are potentially more computationally efficient and can be used to study larger films and longer drying times than particle-based simulations. However, a good approximation for the particle flux resulting from density gradients is required for the continuum model to be accurate. Flux models for drying suspensions have often been chosen empirically, e.g., from phenomenological equations \cite{zhou:prl:2017, reeszimmerman:jfm:2021, howard:jcp:2020}, leading to spirited debate over what is required to describe experimental observations, particularly for multicomponent suspensions \cite{sear:pre:2017,reeszimmerman:jfm:2021, statt:jcp:2018, tang:jcp:2019, howard:jcp:2020}.

Dynamic density functional theory (DDFT) is a convenient framework to systematically construct continuum models for colloidal suspensions from knowledge of the particle-level interactions and dynamics using statistical mechanical principles \cite{marconi:jcp:1999,archer:jcp:2004,archer:jpcm:2005,tevrugt:advphys:2020}. Howard et al.~proposed using DDFT to model small-on-top stratification in drying binary colloidal suspensions \cite{howard:lng:2017,howard:lng:2017b}, and He et al.~recently carried out a comprehensive numerical study of stratification using a similar model \cite{he:lng:2021}. However, other continuum models proposed for drying suspensions \cite{zhou:prl:2017} can also be derived using DDFT formalism. To determine the particle flux, DDFT requires two inputs: a free-energy functional encoding the particle interactions and a model for the particle dynamics. Comparison with experiments and particle-based simulations suggests that both inputs may affect the accuracy of DDFT models \cite{sear:pre:2017,Schulz:2018,statt:jcp:2018,howard:jcp:2020,chun:sm:2019,Park:2022}, and it is accordingly critical to identify appropriate choices.

Testing DDFT models for drying colloidal suspensions against experimental measurements is challenging for two reasons. First, experimentally characterizing the time-resolved distribution of colloids throughout a drying film is difficult, and most comparisons between models and experiments have been limited to the final dried state and/or the surface structure \cite{Fortini:2016,carr:2018,liu:jcis:2018,schulz:2021,Schulz:2018}. Further, because both the free-energy functional and particle dynamics contribute to determining the particle flux (and so also the film structure), it is hard to assess each input of the DDFT model in isolation. For this reason, various approximations of the free-energy functional and particle dynamics have been used and justified, largely qualitatively, based on arguments about the particle concentration, size, or drying regime. For example, drying hard-sphere suspensions have been studied using approximations of the free-energy functional that range in complexity from the dilute (ideal-gas) limit up to nonlocal functionals for concentrated suspensions \cite{howard:lng:2017,howard:lng:2017b,he:lng:2021,Routh:2004,reeszimmerman:jfm:2021,Trueman:2012er,zhou:prl:2017,sear:pre:2017}, but it is unclear in which regimes these approximations apply and whether discrepancies with experiments are due to approximations of the free-energy functional, the particle dynamics, or both.

To make progress in addressing this question, here we test several approximate free-energy functionals that have been used to model drying hard-sphere suspensions. We focus on hard spheres because (1) they are useful models for well-dispersed colloidal particles and (2) hard spheres are often used as a basis for thermodynamic perturbation theories of interacting particles \cite{Weeks:1971}. To test the free-energy functional directly, we use particle-based Brownian dynamics (BD) simulations that assume the same underlying particle dynamics as the DDFT model to create a reference point. Any differences between the DDFT model and the BD simulations can then be attributed to approximations in the free-energy functional or underlying DDFT. Although the functionals we test are well established and characterized for equilibrium systems, we are unaware of a prior systematic comparison of this type for nonequilibrium drying, particularly for multicomponent suspensions.

We first describe the model we studied and the methods we used to perform the BD simulations and DDFT calculations (Sec.~\ref{sec:model}). We then compare the DDFT models to the BD simulations, characterizing the error in the DDFT-predicted volume-fraction profiles for one-component suspensions and the extent of small-on-top stratification for two-component suspensions (Sec.~\ref{sec:results}). We conclude with some numerical and computational considerations we noted in making these comparisons (Sec.~\ref{sec:conclude}).

\section{Model and methods}
\label{sec:model}
We studied one-component and two-component suspensions of hard-sphere particles drying into a film at constant temperature $T$. The two-component suspensions consisted of small (S) and big (B) particles with diameters $d_{\rm S}$ and $d_{\rm B}$, respectively, while the one-component suspensions had only small particles. The film height was oriented along the $z$-axis: the film was supported from below by a repulsive wall with the solvent interface initially a height $H_0$ above it. The particle interactions with the solvent interface were modeled by a repulsive harmonic potential \cite{howard:lng:2017,Fortini:2016, pieranski:prl:1980},
\begin{equation}
\beta \psi^{\rm slv}(z_i,t) = \begin{cases}
\displaystyle \frac{\kappa}{2}\left[z_i-H(t)+\frac{d_i}{2}\right]^2,& z_i > H(t)-\dfrac{d_i}{2} \\
0,& {\rm otherwise}
\end{cases},
\label{eq:slv}
\end{equation}
where $z_i$ and $d_i$ are the vertical position and diameter of particle $i$, $H(t) = H_0-v t$ is the position of the interface receding with constant velocity $v$ at time $t$, and $\beta = 1/(k_{\rm B}T)$ with $k_{\rm B}$ being the Boltzmann constant. We used $\kappa = 100\,d_{\rm S}^{-2}$ and the shift $-d_i/2$ to make the particles remain fully immersed in the solvent with a contact angle of roughly $0^\circ$. This model neglects deformation and capillary interactions \cite{tang:pre:2018} at the liquid interface as well as mass-transfer considerations affecting the drying rate \cite{Routh:1998}. To mimic a large film, periodic boundary conditions were applied parallel to the substrate ($xy$-plane), and only a square cross-section with edge length $L$ was explicitly modeled. The initial average volume fraction of particles of type $i$ (S or B) was then $\phi_{i,0} = N_i \pi d_i^3/(6 H_0 L^2)$, where $N_i$ is the number of particles of type $i$. We computed and analyzed the evolution of the one-dimensional volume fraction profiles $\phi_i(z,t)$ in the film for various $\phi_{i,0}$ and $H_0$ using both BD simulations and DDFT calculations. Details of the implementation of these methods are provided next.

\subsection{Brownian dynamics}
\label{sec:model:bd}
We performed BD simulations to establish a reference point for comparing various DDFT models. Hard-sphere interactions were approximated by the repulsive Weeks--Chandler--Andersen potential \cite{Weeks:1971},
\begin{equation}
\beta u(r_{ij}) = \begin{cases}
\displaystyle 4 \left[\left(\frac{d_{ij}}{r_{ij}}\right)^{12}-\left(\frac{d_{ij}}{r_{ij}}\right)^6\right] + 1,& r_{ij} \le 2^{1/6} d_{ij} \\
0,& {\rm otherwise}
\end{cases},
\label{eq:wca}
\end{equation}
where $r_{ij}$ is the distance between the centers of particles $i$ and $j$ and $d_{ij} = (d_i+d_j)/2$ is the arithmetic mean of their diameters. The repulsive substrate was modeled using a Lennard-Jones 9--3 potential
\begin{equation}
\beta \psi^{\rm w}(z_i) = \begin{cases}
\displaystyle \varepsilon_i \Bigg[ \frac{2}{15}\left(\dfrac{d_i^{\rm w}}{z_i}\right)^9 - \left(\frac{d_i^{\rm w}}{z_i}\right)^3  \\
\displaystyle \quad \quad+ \frac{\sqrt{10}}{3} \Bigg] , \quad z_i \le \left(\dfrac{2}{5}\right)^{1/6} d_i^{\rm w} \\
0, \quad {\rm otherwise}
\end{cases},
\label{eq:ljwall}
\end{equation}
where $d_i^{\rm w} = (d_i+d_{\rm S})/2$ is the contact diameter of a sphere with a hard wall of nominal thickness $d_{\rm S}$ centered at $z=0\,d_{\rm S}$, and $\varepsilon_i$ sets the strength of the repulsion. We used $\varepsilon_{\rm S} = 1$ for the small particles and $\varepsilon_{\rm B} = 100$ for the big particles. The larger value of $\varepsilon_{\rm B}$ was required to reduce the big particle penetration into the wall, which increased with $d_{\rm B}$.

The particle dynamics were simulated using a simple free-draining approximation of the hydrodynamic interactions \cite{Allen:2017,ermak:jcp:1978,wani:jcp:2022}. The position $\vv{r}_i$ of particle $i$ was advanced during a time step $\Delta t$ according to
\begin{equation}
\Delta \vv{r}_i = \gamma_i^{-1} \vv{F}_i \Delta t + \Delta \vv{w}_i,
\label{eq:bd}
\end{equation}
where $\gamma_i$ is the hydrodynamic friction coefficient on $i$ (in isolation), $\vv{F}_i$ is the total force acting on $i$, and $\Delta \vv{w}_i$ is a random displacement that is independent for each particle and Gaussian distributed with zero mean $\langle \Delta \vv{w}_i \rangle = \vv{0}$ and covariance $\langle \Delta\vv{w}_i \Delta \vv{w}_i \rangle = 2 k_{\rm B} T \gamma_i^{-1} \Delta t \vv{I}$, where $\vv{I}$ is the identity tensor. We set the friction coefficients of the small and big particles proportional to their diameters, $\gamma_{\rm B}/\gamma_{\rm S} = d_{\rm B}/d_{\rm S}$, consistent with the low-Reynolds-number drag on a sphere. At infinite dilution, BD gives diffusive motion characterized by the diffusion constant $D_i = k_{\rm B} T/\gamma_i$. All quantities will be reported using energy units based on the thermal energy $k_{\rm B} T$, length units based on the diameter of the small particles $d_{\rm S}$, and time units based on the characteristic time $\tau_{\rm S} = d_{\rm S}^2/D_{\rm S}$ for a small particle to diffuse its own diameter.

Equilibrated initial configurations were prepared by randomly placing the particles into the film without overlap, simulating for $1000\,\tau_{\rm S}$, then recording 5 configurations every $10\,\tau_{\rm S}$. For the one-component suspensions, we fixed $N_{\rm S}$ and varied $L$ to obtain a given nominal $\phi_{0,{\rm S}}$ for films with $H_0 = 50\,d_{\rm S}$ and $100\,d_{\rm S}$ (Table \ref{tab:initone}). (Due to an error computing $N_{\rm S}$, the actual $\phi_{0,{\rm S}}$ is close to but slightly different from its nominal value; we will report all results using the nominal value.) For the two-component suspension, we fixed $L = 72\,d_{\rm S}$, $H_0 = 100\,d_{\rm S}$, and the total initial volume fraction, and we varied $N_{\rm S}$ and $N_{\rm B}$ to obtain a given initial average composition of small and big particles. We studied three different compositions: $(\phi_{0,{\rm S}},\phi_{0,{\rm B}}) = (0.05,0.15)$, $(0.10,0.10)$, and $(0.15,0.05)$. 

Production drying simulations were then initialized from the equilibrated configurations. We computed the one-dimensional particle number-density profiles $\rho_i(z,t)$ at equal time intervals $d_{\rm S}/v$ (equal displacements of the solvent interface) using histograms with bin spacing $0.2\,d_{\rm S}$. The number-density profiles were converted to volume-fraction profiles $\phi_i(z,t)$ by convolving $\rho_i(z,t)$ with the volume of a sphere and averaging across all 5 independent realizations. The simulations were performed using HOOMD-blue \cite{anderson:cms:2020,howard:cpc:2016,howard:cms:2019} (version 2.9.6) with features extended using azplugins \cite{azplugins} (version 0.10.1) and timestep $\Delta t = 10^{-5}\,\tau_{\rm S}$.

\begin{table}
\caption{Model parameters for one-component suspensions.}
\centering
\begin{tabular}{cccccc}
nominal $\phi_{0,{\rm S}}$ & $L$ ($d_{\rm S}$) & $H_0$ ($d_{\rm S}$)& $N_{\rm S}$ & actual $\phi_{0,{\rm S}}$ & $t_{\rm f}$ ($\tau$) \\
\hline
\multirow{2}{*}{0.05} & \multirow{2}{*}{141} & 50 & 93026 & 0.0490 & 225 \\
& & 100 & 187951 & 0.0495 & 900 \\
\multirow{2}{*}{0.10} & \multirow{2}{*}{100} & 50 & 93583 & 0.098 & 200 \\
& & 100 & 189076 & 0.099 & 800 \\
\multirow{2}{*}{0.20} & \multirow{2}{*}{71} & 50 & 94350 & 0.196 & 150\\
& & 100 & 190626 & 0.198 & 600 \\
\multirow{2}{*}{0.40} & \multirow{2}{*}{50} & 50 & 93583 & 0.392 & 50\\
& & 100 & 189076 & 0.396 & 200
\end{tabular}
\label{tab:initone}
\end{table}

The key approximations of BD are that the particle inertia relaxes quickly compared to the typical times for particle diffusion, and solvent-mediated interactions between particles can be neglected \cite{ermak:jcp:1978}. The former approximation is reasonable for many experiments; however, the latter simplification is believed to cause significant discrepancies between simulations and experiments of drying films, particularly for mixtures of particles with different sizes \cite{sear:pre:2017,howard:jcp:2018,statt:jcp:2018,howard:jcp:2020, chun:sm:2019,Park:2022}. Nonetheless, the BD model is well suited for assessing thermodynamic approximations of the DDFT models: the DDFT models will be constructed to have exactly the same particle dynamics as BD so that any deviations between the two must be due to approximations in DDFT.

\subsection{Dynamic density functional theory}
\label{sec:model:ddft}
We tested the accuracy of approximate DDFT calculations \cite{marconi:jcp:1999, archer:jcp:2004, archer:jpcm:2005} against the BD simulations. DDFT models the average number-density profile $\rho_i(\vv{r})$ for each type $i$ at a position $\vv{r}$ rather than instantaneous particle configurations. The Helmholtz free energy $A$ is written as a functional of all densities $\{\rho_i\}$,
\begin{equation}
A[\{\rho_i\}] = A^{\rm ig}[\{\rho_i\}] + A^{\rm ex}[\{\rho_i\}] + \sum_i \int \d{\vv{r}} \rho_i(\vv{r}) \psi_i(\vv{r}).
\end{equation}
The first term $A^{\rm ig}$ is the free energy of an ideal gas,
\begin{equation}
\beta A^{\rm ig}[\{\rho_i\}] = \sum_i \int\d{\vv{r}} \rho_i(\vv{r}) \left(\ln\left[\lambda_i^3 \rho_i(\vv{r})\right]-1\right),
\end{equation}
where $\lambda_i$ is the thermal wavelength of type $i$ and accounts for integrals over the momenta. (The value of $\lambda_i$ will not affect our calculations, so we nominally set its value to $\lambda_i = 1\,d_{\rm S}$ for all types.) The second term $A^{\rm ex}$ is the excess free-energy functional that encodes the particle interactions. The third term accounts for the total external potential $\psi_i$ acting on type $i$.

The equilibrium number-density profiles minimize the grand potential \cite{Evans:1979,henderson:1992},
\begin{equation}
\Omega[\{\rho_i\}] = A[\{\rho_i\}] - \sum_i \int \d{\vv{r}} \rho_i(\vv{r}) \mu_i,
\end{equation}
at constant chemical potential $\mu_i$, or equivalently, must satisfy
\begin{equation}
\frac{\delta \Omega}{\delta \rho_i(\vv{r})} = \frac{\delta A}{\delta \rho_i(\vv{r})} - \mu_i = 0.
\label{eq:dfteq}
\end{equation}
The chemical potential $\mu_i$ can be chosen so that a certain average number of particles $\langle N_i \rangle = \int \d{\vv{r}} \rho_i(\vv{r})$ is obtained.

The nonequilibrium (time-dependent) number-density profiles obey a conservation equation \cite{marconi:jcp:1999,archer:jcp:2004}
\begin{equation}
\frac{\partial \rho_i(\vv{r},t)}{\partial t} + \nabla \cdot \vv{j}_i(\vv{r},t) = 0,
\label{eq:ddft}
\end{equation}
where $\vv{j}_i(\vv{r},t)$ is the total flux of type $i$ at position $\vv{r}$ and time $t$. In standard DDFT \cite{marconi:jcp:1999,archer:jcp:2004,archer:jpcm:2005}, the flux is approximated as
\begin{equation}
\vv{j}_i(\vv{r},t) = -\gamma_i^{-1} \rho_i(\vv{r},t) \nabla\left[\frac{\delta A}{\delta \rho_i(\vv{r},t)}\right].
\label{eq:ddftflux}
\end{equation}
The key assumptions are that the particle dynamics are given by Eq.~\eqref{eq:bd} and that particle correlations out of equilibrium are equivalent to those in the equilibrium system with the same number-density profiles \cite{archer:jcp:2004}.

An accurate excess free-energy functional $A^{\rm ex}$ is critical for DDFT, as it describes thermodynamics due to particle interactions. This functional must usually be approximated because exact solutions are limited to a handful of special cases. The local density approximation (LDA),
\begin{equation}
A^{\rm ex}[\{\rho_i\}] \approx \int \d{\vv{r}} a^{\rm ex}(\{\rho_i(\vv{r})\}),
\end{equation}
treats the inhomogeneous system as a collection of locally homogeneous systems with excess Helmholtz free-energy density $a^{\rm ex}$. The free-energy density can be specified from an equation of state by thermodynamic integration,
\begin{equation}
\beta a^{\rm ex}(\{\rho_i\}) = \rho \int_0^1 \d\lambda \frac{Z(\{\lambda \rho_i\})-1}{\lambda},
\end{equation}
where $\rho = \sum_i \rho_i$ is the total number density and $Z = \beta P/\rho$ is the compressibility factor corresponding to the pressure $P$ \cite{hansen:2006}. We will consider three equations of state for $Z$, which we will discuss in detail in Sec.~\ref{sec:results}.

The LDA is reasonable for nearly homogeneous systems (small concentration gradients and far from interfaces); however, it can fail to capture important physical behaviors of inhomogeneous systems because $a^{\rm ex}$ is only a function of the local density at $\vv{r}$. For example, the LDA does not predict the well-known structuring of a hard-sphere fluid near a hard wall \cite{snook:jcp:1978}. Free-energy functionals of weighted densities \cite{Tarazona:1985}, computed from the number densities at multiple positions, are required to overcome this limitation. For hard spheres, Rosenfeld's fundamental measure theory (FMT) provides highly accurate predictions of equilibrium hard-sphere densities \cite{Rosenfeld:1989uh,Roth:2002vs,Roth:2010ei,hansen:2006}, albeit at the cost of greater mathematical and numerical complexity.

In FMT, the excess free energy is approximated as\cite{Rosenfeld:1989uh}
\begin{equation}
A^{\rm ex}[\{\rho_i\}] \approx \int \d{\vv{r}} \Phi(\{n_\alpha(\vv{r})\}),
\end{equation}
where $\Phi$ is a free-energy density
\begin{align}
\beta \Phi(\{n_\alpha\}) = &-n_0 \ln(1-n_3) + \frac{n_1 n_2 - \vv{n}_1\cdot\vv{n}_2}{1-n_3} \nonumber \\
&+ \frac{n_2^3-3 n_2 (\vv{n}_2\cdot\vv{n}_2)}{24\pi(1-n_3)^2}
\end{align}
that is a function of four weighted scalar densities $n_0$, $n_1$, $n_2$, and $n_3$ and two weighted vector densities $\vv{n}_1$ and $\vv{n}_2$, which we will refer to collectively as $\{n_\alpha\}$. The weighted densities $\{n_\alpha\}$ are obtained by convolving $\{\rho_i\}$ with a corresponding set of weight functions $\{w_\alpha^{(i)}\}$ representing fundamental geometric measures of a sphere of type $i$,
\begin{equation}
n_\alpha(\vv{r}) = \sum_i \int \d{\vv{r}'} \rho_i(\vv{r}') w_\alpha^{(i)}(\vv{r}'-\vv{r}),
\end{equation}
where $n_\alpha$ and $w_\alpha^{(i)}$ are both either scalar or vector. For example, $w_3^{(i)}(\vv{x}) = \theta(d_i/2-|\vv{x}|)$ is the Heaviside function that represents the volume of a sphere of diameter $d_i$, so $n_3(\vv{r})$ is the total volume fraction at $\vv{r}$. The weighted densities can be efficiently evaluated using the convolution theorem with analytical Fourier transforms of $\{w_\alpha^{(i)}\}$ and fast Fourier transforms of $\{\rho_i\}$ \cite{sears:jcp:2003,knepley:2010}. Complete details of the weight functions are available in the literature\cite{Rosenfeld:1989uh} and are excellently reviewed by Roth \cite{Roth:2010ei}.

To implement the DDFT model, we assumed that there were no significant density variations parallel to the substrate ($xy$-plane) due to symmetry, so we represented the number-density profiles as one-dimensional functions $\rho_i(z,t)$ using a regular mesh with spacing of approximately $0.05\,d_{\rm S}$ and densities defined at the centers of the mesh volumes. We used $d_i$ as the hard-sphere diameters and $\psi^{\rm slv}$ [Eq.~\eqref{eq:slv}] for the interactions with the solvent interface, as in the BD simulations. For numerical convenience, we represented the substrate by a true hard wall
\begin{equation}
\beta \psi^{\rm hw}(z) = \begin{cases}
\infty,& z < z^{\rm hw}+d_i/2 \\
0,& {\rm otherwise}
\end{cases}.
\label{eq:hardwall}
\end{equation}
We empirically selected $z^{\rm hw} = 0.25\,d_{\rm S}$ to match the effective exclusion of $\psi^{\rm w}$ [Eq.~\eqref{eq:ljwall}] for the small particles. We also confined the densities using another hard wall placed at one largest-particle diameter above the initial solvent interface and with opposite orientation, but this detail does not materially affect our calculations because the density above the solvent interface was essentially zero.

We initialized the particle densities uniformly between the hard walls based on $\phi_{0,{\rm S}}$, then used fixed-point iteration to solve for the initial equilibrium profiles. At the $k$-th iteration, a trial density profile was computed, up to a proportionality constant, from Eq.~\eqref{eq:dfteq},
\begin{equation}
\rho_i^{(k*)} \propto \exp\left[-\beta\left(\frac{\delta A^{\rm ex}}{\delta \rho_i^{(k)}} + \psi_i \right)\right],
\end{equation}
using the densities at the $k$-th iteration $\{\rho_i^{(k)}\}$. We then normalized $\rho_i^{(k*)}$ to ensure the average number of particles $\langle N_i \rangle$ remained constant and equivalent to the BD simulations. The density profile at the next iteration was computed by mixing $\{\rho_i^{(k*)}\}$ with $\{\rho_i^{(k)}\}$
\begin{equation}
\rho_i^{(k+1)} = \alpha \rho_i^{(k*)} + (1-\alpha)\rho_i^{(k)}
\end{equation}
with mixing parameter $\alpha = 10^{-4}$. The calculation was determined to be numerically converged when all densities changed by less than $10^{-8}\,d_{\rm S}^{-3}$ between iterations in the one-component calculations and $10^{-10}\,d_{\rm S}^{-3}$ in the two-component calculations.

We then simulated drying by solving Eq.~\eqref{eq:ddft} with a finite-volume scheme. The one-dimensional flux $j_i(z,t)$ at the faces of each mesh volume was determined from Eq.~\eqref{eq:ddftflux}. The gradient of $\delta A^{\rm ig}/\delta \rho_i$ was evaluated analytically to give the usual Fick's law diffusion, while gradients of the other functional derivatives were evaluated numerically. Gradients of $\rho_i$, $\delta A^{\rm ex}/\delta \rho_i$, and $\psi_i$ at the mesh faces were evaluated by differencing their values at the mesh centers, and the density at the mesh faces was linearly interpolated. We enforced that there was no flux across faces with hard walls. The finite-volume scheme guarantees that the number of particles is exactly conserved, as fluxes between all mesh volumes balance. For numerical stability, we used the implicit Euler method \cite{Press:2002} to integrate Eq.~\eqref{eq:ddft} with time step $\Delta t = 10^{-5}\,\tau_{\rm S}$ and fixed-point iteration with mixing parameter $\alpha = 0.5$ and tolerance of $10^{-8}\,d_{\rm S}^{-3}$ to solve for $\rho_i(z,t+\Delta t)$ from $\rho_i(z,t)$. We found that integration using the explicit Euler method often led to numerical instability or required an unacceptably small time step. We recorded $\rho_i(z,t)$ and computed $\phi_i(z,t)$ using the same procedure as in the BD simulations. Even with the implicit Euler time-integration scheme, some of the DDFT calculations still failed to generate density profiles after a certain time due to numerical instability. We have chosen not to show any results when a calculation did not run to completion, and we have noted these failed calculations in Sec.~\ref{sec:results}.

\section{Results and Discussion}
\label{sec:results}
\subsection{One-component suspension}
We first performed BD simulations of the one-component suspensions at ``fast'' drying rates that produced concentration gradients. We fixed the P\'{e}clet number ${\rm Pe}_{\rm S}$ of the small particles, defined using their diffusion constant $D_{\rm S}$ at infinite dilution, at ${\rm Pe}_{\rm S} = 10$ based on typical values in experiments \cite{Schulz:2018}. This choice gave $v = 0.2\,d_{\rm S}/\tau$ for $H_0 = 50\,d_{\rm S}$ and $v = 0.1\,d_{\rm S}/\tau$ for $H_0 = 100\,d_{\rm S}$. We dried the films until they reached a final average volume fraction of roughly $\phi_{{\rm f},{\rm S}} = 0.5$, meaning that each simulation ran for total time $t_{\rm f} = (H_0/v)(1-\phi_{0,{\rm S}}/\phi_{{\rm f},{\rm S}})$. We chose to stop our simulations at this point because the assumptions underlying our model become less valid as close-packing is approached.

The volume fraction profile $\phi_{\rm S}(z,t)$ evolved in a qualitatively similar way for the different initial volume fractions and film heights studied, so we will highlight only the result for $\phi_{0,{\rm S}} = 0.10$ and $H_0 = 100\,d_{\rm S}$ here [Fig.~\ref{fig:single_time}(a)]. Data for the other initial conditions is included in the Supplementary Material (Figs.~S1--S7). The particles were initially nearly uniformly distributed in the film except close to the solvent interface and the substrate, where some concentration enhancement was observed. This thermodynamic effect is well known for hard spheres against hard surfaces \cite{snook:jcp:1978} and becomes more pronounced at larger volume fractions. The increased concentration near the surface eventually leads to particle-scale layering that appears in $\phi_{\rm S}(z,t)$ as oscillations with characteristic spacing $d_{\rm S}$. We observed all of these behaviors in our equilibrated initial configurations, as expected.
\begin{figure}
    \centering
    \includegraphics{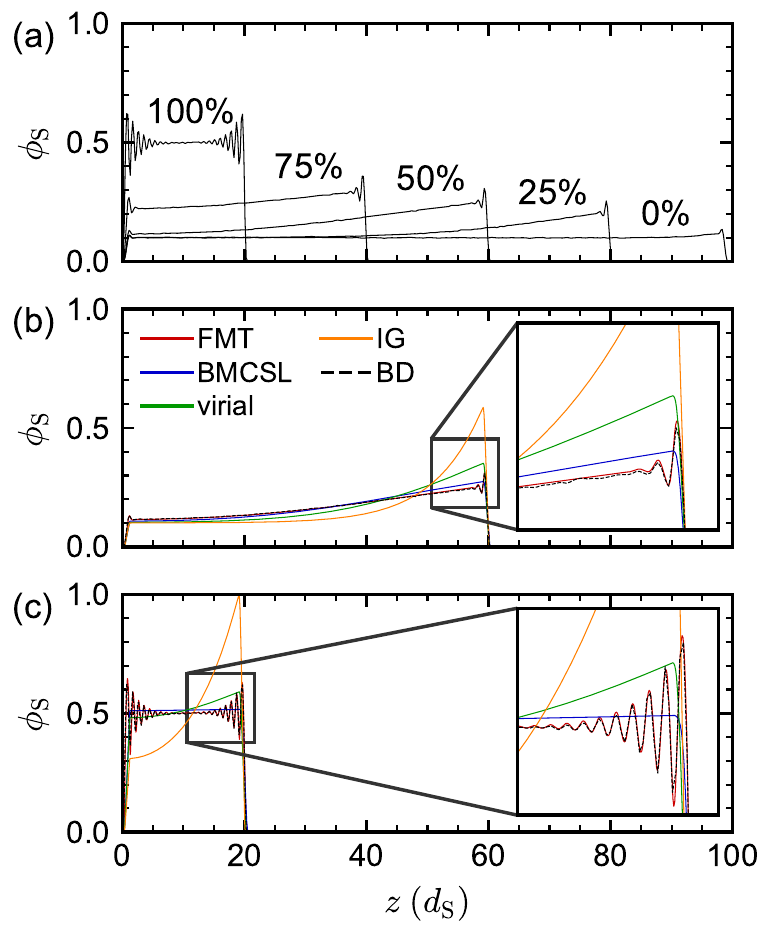}
    \caption{Small-particle volume fraction profile $\phi_{\rm S}$ in drying one-component suspension with $\phi_{0,{\rm S}} = 0.1$, $H_0 = 100\,d_{\rm S}$, and $v = 0.1\,d_{\rm S}/\tau$. (a) BD simulation at times corresponding to 0\%, 25\%, 50\%, 75\% and 100\% of the total drying time $t_{\rm f}$. (b)--(c) Comparison of DDFT calculations with different free-energy functionals to BD simulation at (b) 50\% of $t_{\rm f}$ and (c) 100\% of $t_{\rm f}$.}
    \label{fig:single_time}
\end{figure}

Once drying began, a time-dependent concentration gradient developed because the particles were collected by the moving solvent interface faster than they diffused away (${\rm Pe}_{\rm S} \gg 1$). The gradient initially formed near the top of the film [25\% of total drying time $t_{\rm f}$ in Fig.~\ref{fig:single_time}(a)] then propagated toward the bottom [50\% of $t_{\rm f}$ in Fig.~\ref{fig:single_time}(a)]. At later times, the concentration gradient across the film tended to decrease as the film densified [75\% of $t_{\rm f}$ in Fig.~\ref{fig:single_time}(a)]. By the end of the simulation, there was no longer a substantial concentration gradient but $\phi_{\rm S}$ showed oscillations due to layering. We visually inspected the final configurations and did not identify any significant crystalline order in these layers. The most oscillations in $\phi_{\rm S}$ at the end of the simulations, characterized by spatial extent relative to the final film height, were observed for the smallest initial volume fraction ($\phi_{0,{\rm S}} = 0.05$). This is because a thinner final film height (more drying) was needed to achieve the final target volume fraction of $\phi_{{\rm f},{\rm S}} = 0.5$ when $\phi_{0,{\rm S}}$ was smaller, and the particles were accordingly more confined in this final state. In contrast, films with larger $\phi_{0,{\rm S}}$ dried less and had a larger bulk region.

We noted that $\phi_{\rm S}(z,t)$ for $H_0 = 50\,d_{\rm S}$ nearly collapsed with that for $H_0 = 100\,d_{\rm S}$ when $z$ was rescaled by $H_0$ and $t$ was rescaled by $H_0/v$ (Fig.~S8). The primary differences were in the interfacial oscillations, which are characterized by the particle diameter rather than the film height and so do not collapse when scaled by $H_0$. The volume-fraction profiles did not collapse when $v$ was held constant instead of ${\rm Pe}_{\rm S}$. This supports the film P\'{e}clet number as an appropriate dimensionless number for this advection--diffusion process.

After establishing our reference BD simulations, we performed DDFT calculations with different free-energy functionals based on either a local-density approximation using a bulk equation of state (EOS) or a weighted-density approximation using FMT \cite{hansen:2006}. We tested three different EOSs for the LDA. The simplest was the ideal-gas (IG) EOS for which $Z=1$ and hence $\beta a^{\rm ex} = 0$. The IG EOS has been used, at least partially \cite{sear:pre:2017,Trueman:2012er,reeszimmerman:jfm:2021}, in some drying studies but does not include any interparticle interactions. As such, it is most valid in very dilute suspensions where particles rarely interact and becomes less reasonable as the film densifies.

To incorporate particle interactions at finite concentrations, we first considered a second-order virial EOS. The virial EOS is a low-density expansion that includes pairwise interactions between particles, and its compressibility factor is
\begin{equation}
Z = 1 + \rho \sum_{i,j} x_i x_j B_{ij},
\label{eq:virial2}
\end{equation}
where $x_i = \rho_i/\rho$ is the mole fraction of type $i$, the indexes $i$ and $j$ of the double sum run over all types, and $B_{ij} = (2\pi/3)(d_i/2+d_j/2)^3$ for a hard-sphere mixture \cite{rigby:1963}. For the one-component suspension, Eq.~\eqref{eq:virial2} simplifies to
\begin{equation}
Z = 1 + 4 \phi_{\rm S}.
\label{eq:virial1}
\end{equation}
The virial EOS has often been used to study drying or sedimentation, perhaps in part because it has a simple functional form that lends itself to analytical treatment. For example, the virial EOS forms the basis for Batchelor's classic calculations of diffusion in hard-sphere suspensions \cite{batchelor:jfm:1976,batchelor:jfm:1983}. The virial EOS was also used by Zhou, Jiang, and Doi to demonstrate the importance of cross-interactions between components in the stratification of hard-sphere mixtures \cite{zhou:prl:2017}. However, it has been questioned whether the virial EOS is sufficiently accurate at the higher concentrations that are typical of experiments and inevitably encountered during drying \cite{Schulz:2018}.

We accordingly also tested the Boubl\'{i}k--Mansoori--Carnahan--Starling--Leland (BMCSL) EOS, which has been demonstrated to be a highly accurate model for bulk hard-sphere mixtures across a range of concentrations \cite{boublik:1970,Mansoori:1971vt}. Indeed, one of us previously used the BMCSL EOS to model drying mixtures of hard-sphere polymers with good success \cite{howard:lng:2017b}. The compressibility factor for the BMCSL EOS is
\begin{equation}
Z = \frac{6}{\pi\rho}\left[\frac{\xi_0}{1-\xi_3}+\frac{3\xi_1\xi_2}{(1-\xi_3)^2}+\frac{(3-\xi_3)\xi_2^3}{(1-\xi_3)^2}\right],
\label{eq:bmcsl2}
\end{equation}
where $\xi_m$ = $\sum_i \rho_i \pi d_i^m/6$. For the one-component suspension, Eq.~\eqref{eq:bmcsl2} simplifies to the Carnahan--Starling EOS \cite{Carnahan:1969bg},
\begin{equation}
Z = \frac{1 + \phi_{\rm S} + \phi_{\rm S}^2 - \phi_{\rm S}^3}{(1-\phi_{\rm S})^3}.
\label{eq:bmcsl1}
\end{equation}
The compressibility factor for the BMCSL EOS approaches that of the virial EOS at low densities but predicts a stronger dependence on concentration at higher densities, ultimately diverging when $\xi_3 = 1$ for the mixture or when $\phi_{\rm S} = 1$ for the one-component suspension. Both the IG and virial EOSs do not capture this divergence.

In fact, though, $Z$ should diverge even earlier as the fluid approaches random close packing \cite{torquato:2000}, which occurs at $\phi_{\rm S} \approx 0.64$ for the one-component suspension. To capture this behavior, Routh and coworkers have used variations of an empirical EOS \cite{Routh:1998,Routh:2004},
\begin{equation}
Z = \frac{Z_0}{0.64-\phi_{\rm S}}
\label{eq:Routh}
\end{equation}
where $Z_0$ is a prefactor, to model drying one-component suspensions and a similar form incorporating both $\phi_{\rm S}$ and $\phi_{\rm B}$ to model two-component suspensions \cite{Trueman:2012er,reeszimmerman:jfm:2021}. We note, however, that Eq.~\eqref{eq:Routh} does not have the appropriate limiting behavior as $\phi_{\rm S} \to 0$. Eq.~\eqref{eq:Routh} requires $Z_0 = 0.64$ to recover the ideal-gas limit $Z \to 1$ as $\phi_{\rm S} \to 0$, but this value of $Z_0$ does not yield the correct second virial coefficient [Eq.~\eqref{eq:virial1}] after series expansion for small $\phi_{\rm S}$. Any other choice of $Z_0$ (e.g. \cite{Routh:2004}, $Z_0=1$) results in a divergent integral for $\beta a^{\rm ex}$. These issues can be circumvented by applying Eq.~\eqref{eq:Routh} only at high $\phi_{\rm S}$ (rather than all $\phi_{\rm S}$) as a patch to the BMCSL EOS \cite{russel:1989}. For example, a continuous function $Z(\phi_{\rm S})$ that diverges when $\phi_{\rm S} \to 0.64$ is obtained \cite{russel:1989} by using Eq.~\eqref{eq:bmcsl1} when $\phi_{\rm S} \le 0.5$ and Eq.~\eqref{eq:Routh} with $Z_0 = 1.82$ when $\phi_{\rm S} > 0.5$. However, it is unclear if such a patch is even needed to model drying because diffusion can prevent regions of high concentration from forming until late in the process. Moreover, the one-component hard-sphere fluid is only metastable with respect to the hard-sphere crystal above $\phi_{\rm S} \approx 0.5$ \cite{russel:1989}, but our DDFT model cannot capture this freezing transition with the assumed symmetry or free-energy functionals \cite{Rosenfeld:1989uh,Roth:2010ei}. Accordingly, we have chosen not to use Eq.~\eqref{eq:Routh} as its own EOS or as a correction to Eq.~\eqref{eq:bmcsl1}.

We performed DDFT calculations using these three LDA functionals for the same $\phi_{0,{\rm S}}$, $H_0$, and ${\rm Pe}_{\rm S}$ as the BD simulations, and we compared $\phi_{\rm S}$ at the same extents of drying. As we expected, the BMCSL EOS was the most accurate and the IG EOS was the least accurate of the three LDA functionals, both at intermediate [Fig.~\ref{fig:single_time}(b)] and final [Fig.~\ref{fig:single_time}(c)] times. In particular, both the IG and virial EOSs predicted concentrations at the top of the film (and concentration gradients) that were larger than in the BD simulations. The BMCSL EOS gave $\phi_{\rm S}$ that was initially similar to the BD simulations, but discrepancies became apparent at the later stages of drying as oscillations in $\phi_{\rm S}$ developed in the more concentrated films. These oscillations cannot be captured by the LDA functionals we considered, so we performed the same DDFT calculations using a weighted-density approximation based on FMT. FMT is known to be a highly accurate free-energy functional for confined hard-sphere fluids at equilibrium \cite{Roth:2010ei} but is more costly to evaluate than the LDA functionals. As expected, FMT predicted an initial equilibrium volume-fraction profile that was in excellent agreement with the BD simulations. Importantly, though, FMT also faithfully reproduced $\phi_{\rm S}$ throughout the nonequilibrium drying process.

\begin{figure}
    \centering
    \includegraphics{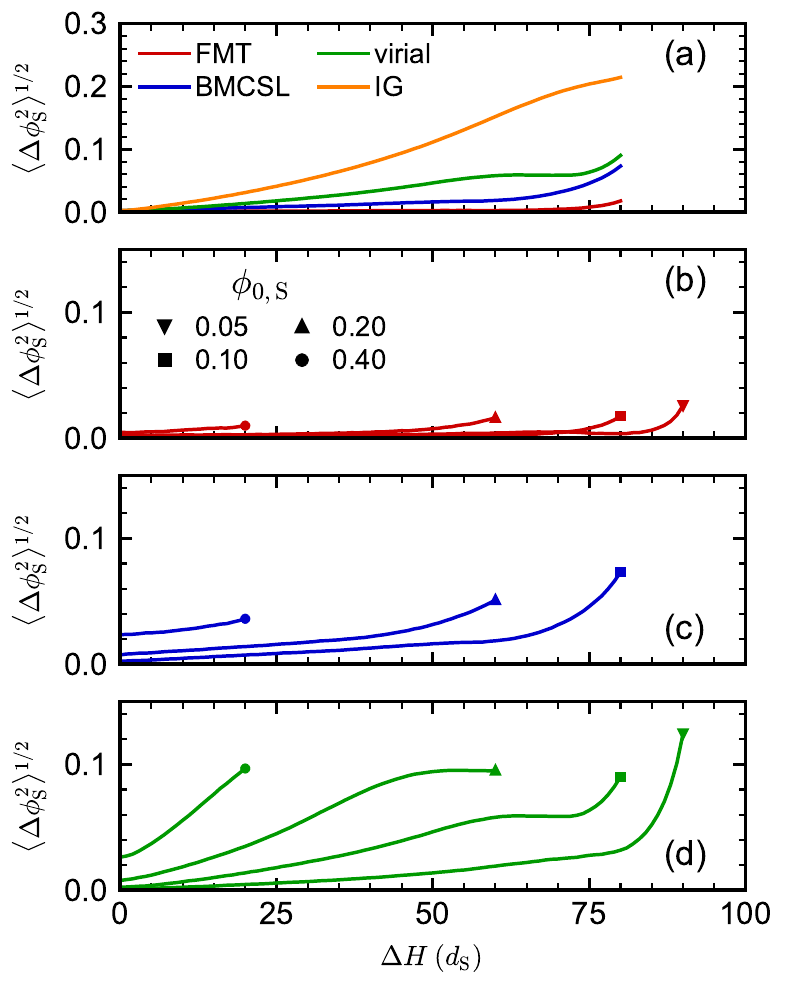}
    \caption{Root mean squared error $\langle \Delta \phi_{\rm S} \rangle^{1/2}$ in DDFT calculations of small-particle volume fraction profiles compared to BD simulations as a function of solvent interface displacement $\Delta H = H_0-H(t)$  when $H_0 = 100\,d_{\rm S}$ and $v = 0.1\,d_{\rm S}/\tau$. (a) Comparison of error for the various free-energy functionals for $\phi_{0,{\rm S}} = 0.1$, as in Fig.~\ref{fig:single_time}. (b)--(d) Error as initial volume fraction $\phi_{0,{\rm S}}$ is varied for (b) FMT and LDA using the (c) BMCSL EOS and (d) virial EOS. Note that the result for $\phi_{0,{\rm S}} = 0.05$ is not included in (c) because of numerical failure at the penultimate displacement. Figure S9 shows the same for $H_0 = 50\,d_{\rm S}$ and $v=0.2\,d_{\rm S}/\tau$.}
    \label{fig:single_error}
\end{figure}

To quantify the accuracy of the free-energy functionals, we computed the instantaneous mean squared error in $\phi_{\rm S}$ between the DDFT calculations and the BD simulations,
\begin{equation}
\langle \Delta \phi_{\rm S}^2(t) \rangle = \frac{1}{H(t)} \int_{0}^{H(t)} \d{z} [\phi_{\rm S}^{\rm DDFT}(z,t)-\phi_{\rm S}^{\rm BD}(z,t)]^2.
\end{equation}
Because the DDFT calculations and BD simulations determined $\phi_{\rm S}$ on different discrete meshes, we interpolated the DDFT data onto the BD mesh before evaluating the integral. In agreement with our qualitative observations, the root mean squared error $\langle \Delta \phi_{\rm S}^2 \rangle^{1/2}$ corresponding to the conditions of Fig.~\ref{fig:single_time} was largest for the LDA with the IG EOS and lowest for FMT at all times during the calculation [Fig.~\ref{fig:single_error}(a)]. The error typically grew with time for all of the functionals, with the LDA functionals accruing error faster than FMT.

This increase is primarily due to two effects: accumulation of error over time and decreasing accuracy of the LDA functionals at higher concentrations. In particular, the IG EOS is accurate only at low volume fractions because it neglects interparticle interactions, the virial EOS is accurate over a slightly larger range of volume fractions than the IG EOS because it captures dilute interactions, but the BMCSL EOS is valid over a large range of volume fractions below close packing. Accordingly, the BMCSL EOS had the smallest error of the three LDA functionals. This interpretation is corroborated by our calculations at other initial volume fractions [Figs.~\ref{fig:single_error}(c)--(d)]: the virial and BMCSL EOSs maintained a smaller error for longer in the drying process when the suspension was initially more dilute. We also note that there was an increase in the error near the end of drying for all initial conditions. For the LDA functionals, this is partially due to failure to capture oscillations in $\phi_{\rm S}$ in confinement. Another contribution to this error for all functionals, including FMT [Fig.~\ref{fig:single_error}(b)], were differences in the volume fraction profile near the bottom substrate. These differences are due to choice of models for the substrate [Eqs.~\ref{eq:ljwall} and \ref{eq:hardwall}] and are present even at equilibrium; however, they contribute more significantly to the average error when the film height decreases.

\begin{figure}
    \centering
    \includegraphics{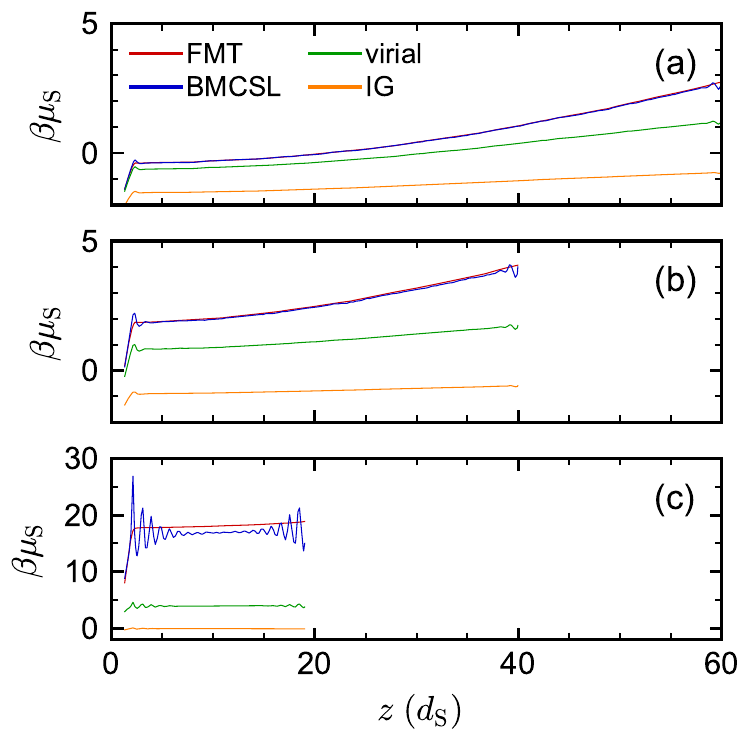}
    \caption{Intrinsic chemical potential $\mu_{\rm S}(z,t)$ predicted by different free-energy functionals for the BD density profiles corresponding to Fig.~\ref{fig:single_time}(a) at (a) 50\%, (b) 75\%, and (c) 100\% of the total drying time $t_{\rm f}$.}
    \label{fig:single_potential}
\end{figure}

Ultimately, the different free-energy functionals produce different time-dependent volume fraction profiles because they give different intrinsic chemical potentials $\mu_i(z,t) = \delta A/\delta \rho_i(z,t)$; gradients of $\mu_i$ drive diffusion [Eq.~\eqref{eq:ddftflux}]. To characterize these differences, we used the BD density profiles as common inputs to DDFT and computed $\mu_{\rm S}(z,t)$ (Fig.~\ref{fig:single_potential}). Given statistical uncertainty in the BD data, we first smoothed the density profiles using a window average of 10 adjacent points, then interpolated onto the DDFT mesh. Among the LDA functionals, the BMCSL EOS gave the largest values of $\mu_{\rm S}$ and its gradient; this makes sense because the compressibility factor (and corresponding bulk chemical potential) of the BMCSL EOS [Eq.~\eqref{eq:bmcsl1}] has a stronger dependence on $\phi_{\rm S}$ than the virial [Eq.~\eqref{eq:virial1}] or IG EOSs. As drying progresses, weaker gradients in $\mu_{\rm S}$ lead to faster accumulation at the solvent interface, consistent with Figs.~\ref{fig:single_time} and \ref{fig:single_error}. Compared to the BMCSL EOS, FMT predicted similar (but slightly larger) values for $\mu_{\rm S}$ during early and intermediate stages of drying [Figs.~\ref{fig:single_potential}(a)--(b)]. This makes sense because Rosenfeld's FMT reduces to the Percus--Yevick EOS \cite{percus:pr:1958,wertheim:prl:1963} for a bulk hard sphere fluid, which is similar (but not identical) to the BMCSL EOS. However, at late stages of drying where density oscillations form, there was significant disagreement between FMT and the BMCSL EOS [Fig.~\ref{fig:single_potential}(c)]. This is expected because, as a LDA functional, the BMCSL EOS cannot support spatial oscillations in density and the variations in $\mu_{\rm S}$ act to eliminate them. This calculation explains why FMT and BMCSL produce similar $\phi_{\rm S}$ profiles when the system is not strongly confined and gradients are not large, but differ more significantly near interfaces and as confinement effects become significant.

Overall, FMT was the best performing functional with a remarkably low maximum root mean squared error of 0.035 for all initial conditions tested, but the BMCSL EOS gave a reasonable approximation prior to confinement. We note that both the FMT functional and the BMCSL EOS do not diverge when the local volume fraction approaches typical values of random close-packing. Yet, our DDFT calculations with FMT were still in excellent agreement with the BD simulations even at late stages of drying when the maximum local volume fraction was close to $\phi_{\rm S} \approx 0.6$. This finding justifies neglect of a correction like Eq.~\eqref{eq:Routh} to the compressibility factor. The virial EOS had acceptable performance for very dilute suspensions; however, it quickly became inaccurate as the concentration increased. The IG EOS vastly overpredicted concentration gradients and yielded unphysical values of $\phi_{\rm S}$. Hence, we recommend caution when applying the virial EOS to nondilute suspensions, and we do not recommend using the IG EOS. Computational considerations help guide selection between the FMT functional and LDA functional with the BMCSL EOS, and we will comment more on this point in Sec.~\ref{sec:conclude}.

\subsection{Two-component suspension}
\begin{figure*}
    \centering
    \includegraphics{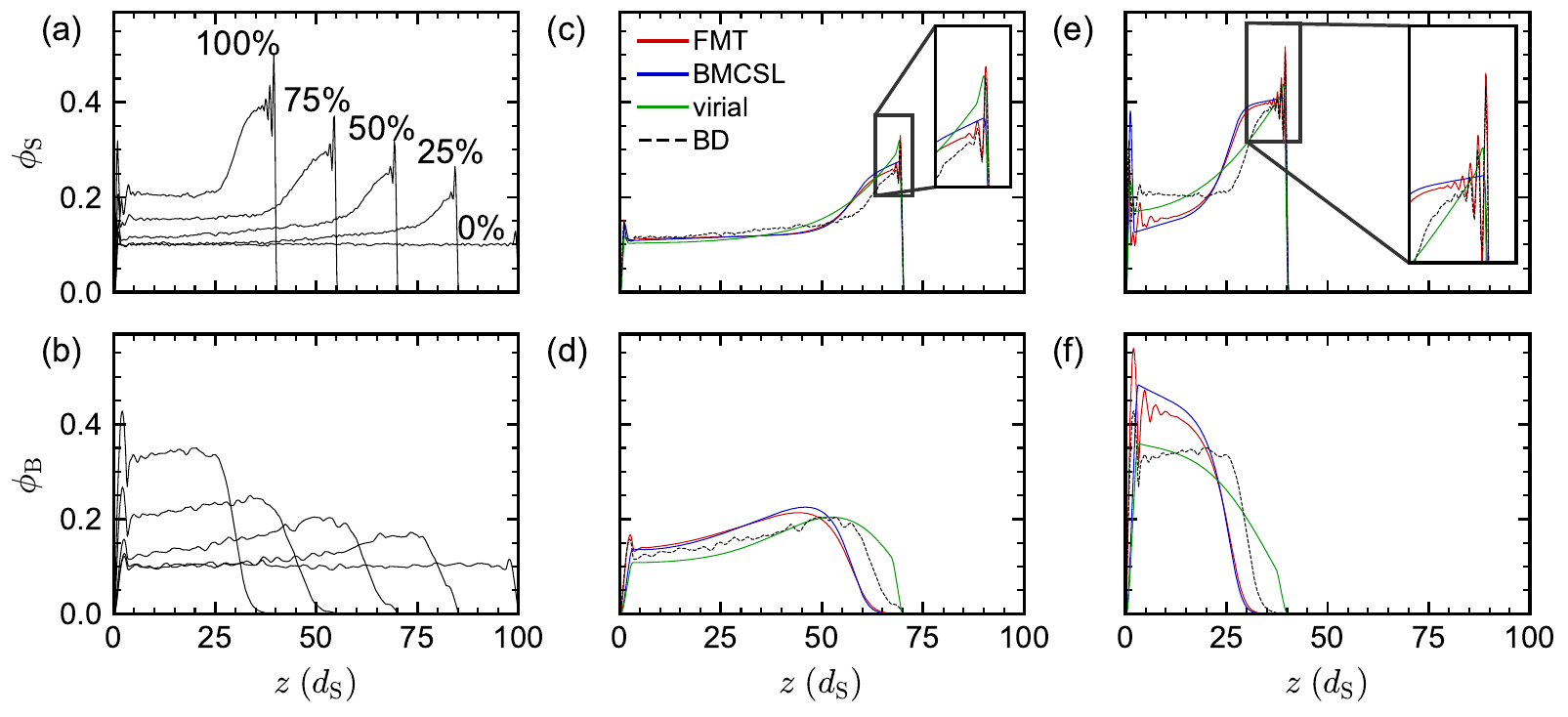}
    \caption{Small-particle volume fraction profile $\phi_{\rm S}$ (top row) and big-particle volume fraction profile $\phi_{\rm B}$ (bottom row) in drying two-component suspension with $d_{\rm B} = 3\,d_{\rm S}$, 1:1 initial composition ($\phi_{0,{\rm S}} = 0.10$, $\phi_{0,{\rm B}} = 0.10$), $H_0 = 100\,d_{\rm S}$, and $v = 0.1\,d_{\rm S}/\tau$. (a)--(b) BD simulation at times corresponding to 0\%, 25\%, 50\%, 75\% and 100\% of the total drying time $t_{\rm f}$. (c)--(f) Comparison of DDFT calculations with different free-energy functionals to BD simulation at (c)--(d) 50\% of $t_{\rm f}$ and (e)--(f) 100\% of $t_{\rm f}$.}
    \label{fig:binary_time}
\end{figure*}
After investigating one-component suspensions of small particles, we proceeded to study two-component suspensions of both small and big particles. We considered films with initial height $H_0 = 100\,d_{\rm S}$ and the same small-particle film P\'{e}clet number ${\rm Pe}_{\rm S} = 10$ as we used for the one-component suspensions, meaning that $v = 0.1\,d_{\rm S}/\tau$. We performed simulations with big-particle diameters $d_{\rm B} = 3\,d_{\rm S}$ and $6\,d_{\rm S}$, corresponding to big-particle film P\'{e}clet numbers ${\rm Pe}_{\rm B} = 30$ and 60, respectively, defined using their diffusion coefficient $D_{\rm B}$. We also varied the relative initial composition of small and big particles, keeping the total initial volume fraction constant at $\phi_{0,{\rm S}} + \phi_{0,{\rm B}} = 0.2$. We will denote the initial composition using the ratio of $\phi_{0,{\rm S}}$ to $\phi_{0,{\rm B}}$, so the three initial compositions we studied were 1:3, 1:1, and 3:1. As for the one-component suspensions, the film was dried for $600\,\tau$ to reach a final average volume fraction of $\phi_{{\rm f},{\rm S}} + \phi_{{\rm f},{\rm B}} = 0.5$.

\begin{figure}
    \centering
    \includegraphics{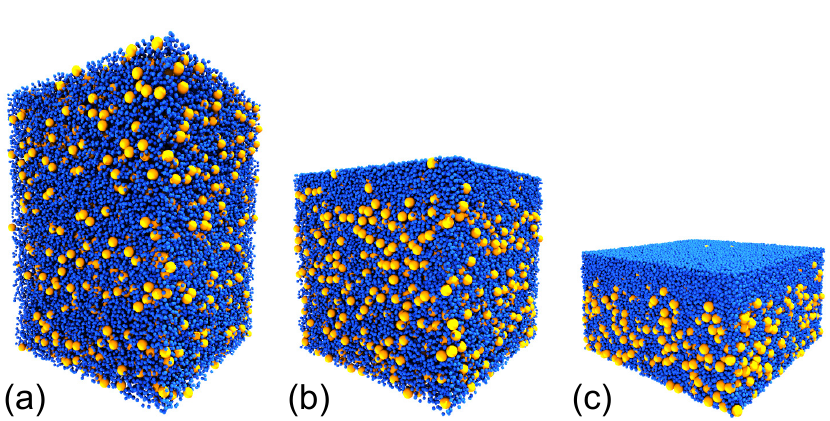}
    \caption{BD trajectory snapshots of small particles (blue) and big particles (orange) for the conditions corresponding to Fig.~\ref{fig:binary_time} at (a) 0\%, (b) 50\%, and (c) 100\% of $t_{\rm f}$. Snapshots rendered using Visual Molecular Dynamics (version 1.9.3) \cite{vmd}.}
    \label{fig:BD}
\end{figure}

Based on prior simulation studies using Langevin dynamics \cite{Fortini:2016,howard:lng:2017}, we expected the drying mixtures to stratify into layers with the small particles on top of the big particles. Our BD simulations were consistent with this expectation for both big-particle diameters and all initial relative compositions studied. Figures~\ref{fig:binary_time}(a)--(b) show the typical time evolution of the volume fraction profiles $\phi_{\rm S}(z,t)$ and $\phi_{\rm B}(z,t)$ in the BD simulations for $d_{\rm B} = 3\,d_{\rm S}$ at the 1:1 initial composition. Data for the other initial compositions (Figs.~S10--S11) and for $d_{\rm B} = 6\,d_{\rm S}$ (Figs.~S13--S15) are available in the Supplementary Material. Initially, both the small and big particles were nearly uniformly distributed in the film except in narrow regions of width comparable to $d_{\rm B}$ near the substrate and interface. During drying, $\phi_{\rm S}$ increased in the top of the film near the solvent interface, like in the one-component suspensions [Fig.~\ref{fig:single_time}(a)], but did not increase substantially in the bottom of the film, unlike in the one-component suspensions. This resulted in a steeper, more localized gradient in $\phi_{\rm S}$ than in the one-component suspensions. Concurrently, $\phi_{\rm B}$ decreased in the top of the film and increased near the bottom, eventually becoming close to zero in the top of the film. This evolution of $\phi_{\rm S}$ and $\phi_{\rm B}$ is consistent with strong ``small-on-top'' stratification into two layers, where the big particles are nearly completely excluded from a layer enriched in small particles at the top of the film (Fig.~\ref{fig:BD}). We again visually inspected both the small-particle top layer and the big particles collected below it for crystallization [Fig.~\ref{fig:BD}(c)], but we did not find any obvious ordering of either during the simulated time interval.

As has been established elsewhere \cite{Fortini:2016,howard:lng:2017,zhou:prl:2017,sear:pre:2017}, stratification is fundamentally driven by interactions between the small and big particles. From a thermodynamic perspective, high concentrations of small particles penalize the presence of big particles, i.e., lead to local regions of high intrinsic chemical potential. Hence, a gradient in $\phi_{\rm S}$ can induce a gradient in $\mu_{\rm B}$ that serves as a driving force for big-particle diffusion and generates a type of phoretic motion \cite{Brady:2011bh,sear:pre:2017,howard:jcp:2020}. Consistent with this idea, we observed the strongest separation of big particles at the 3:1 initial composition, which had the largest amount of small particles. We also noted that, similarly to our prior work \cite{howard:lng:2017}, there was a stronger separation for $d_{\rm B} = 6\,d_{\rm S}$ than for $d_{\rm B} = 3\,d_{\rm S}$. However, we emphasize that the motion in our BD simulations is for free-draining hydrodynamics in an implicit solvent. It has been shown that neglect of solvent-mediated hydrodynamic interactions may lead to overprediction of stratification and incorrect dependence on $d_{\rm B}$ compared to predictions made using a fluid-mechanics analysis accounting for the solvent \cite{sear:pre:2017,statt:jcp:2018,howard:jcp:2020,Park:2022}. Faithfully capturing the role of the solvent in stratification is an important topic but it is not the focus of this work, and the BD simulations remain suitable for testing the DDFT free-energy functionals.

\begin{figure*}
    \centering
    \includegraphics{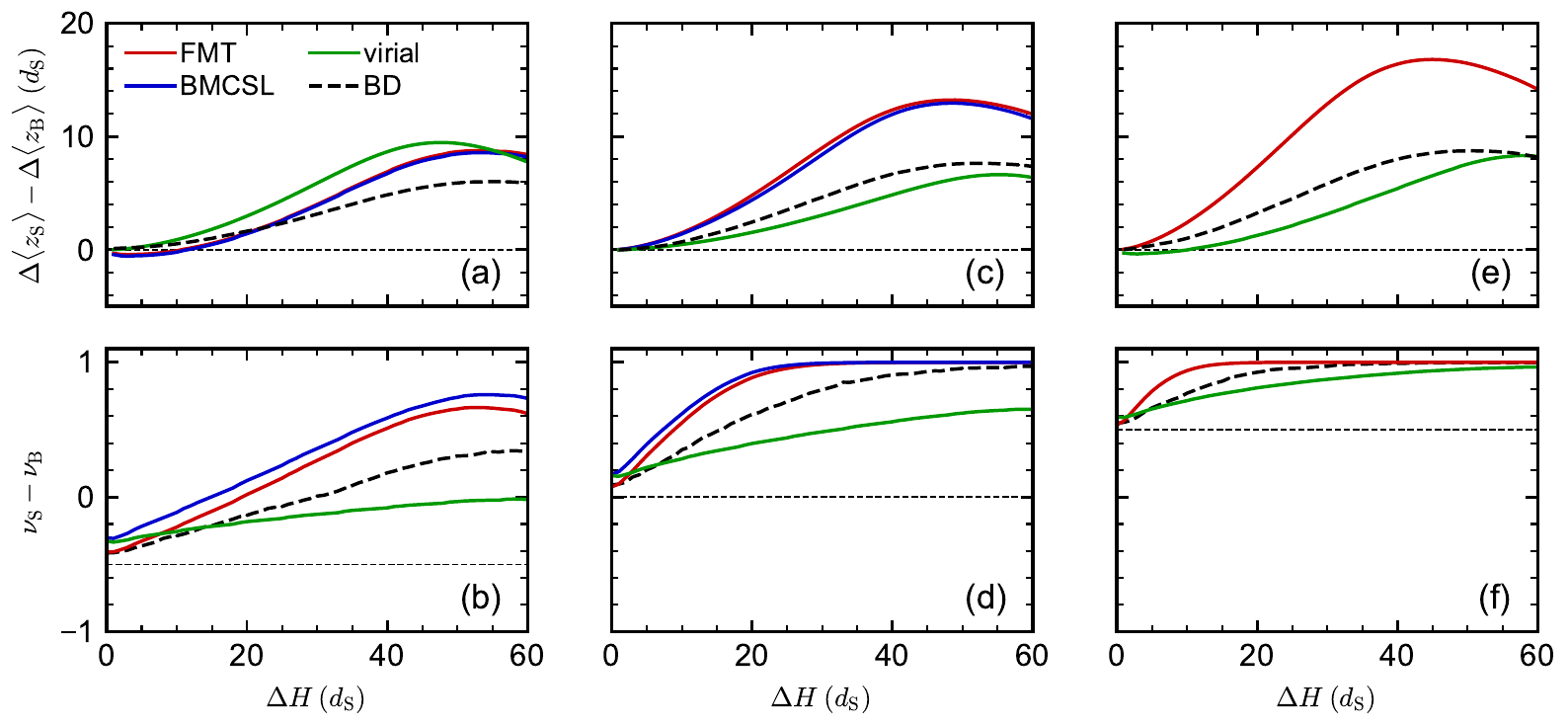}
    \caption{Extent of stratification in drying two-component suspension with $d_{\rm B} = 3\,d_{\rm S}$, $H_0 = 100\,d_{\rm S}$, and $v=0.1\,d_{\rm S}/\tau$ as a function of solvent interface displacement $\Delta H$. Stratification was quantified through both the difference in center-of-mass displacement $\Delta\Delta z = \Delta\langle z_{\rm S}\rangle-\Delta\langle z_{\rm B}\rangle$ using Eq.~\eqref{eq:zcom} (top row) and the difference in surface composition by volume $\Delta \nu = \nu_{\rm S}-\nu_{\rm B}$ using Eq.~\eqref{eq:surfvol} (bottom row). The dotted lines mark the expected values for a homogeneous film starting from the initial compositions (a)--(b) 1:3, (c)--(d) 1:1 (same as Fig.~\ref{fig:binary_time}), and (e)--(f) 3:1. Note that the BMCSL result is not included in (e)--(f) due to numerical failure at roughly 75\% of the total drying time. Figure S16 shows the same for $d_{\rm B} = 6\,d_{\rm S}$.}
    \label{fig:extent_of_strat}
\end{figure*}

We performed DDFT simulations of the two-component suspensions at the same conditions as the BD simulations using LDA functionals with the virial and BMCSL EOSs as well as FMT. We chose not to test the LDA with the IG EOS because an ideal gas does not have interactions between the small and big particles and so cannot predict the stratification observed in the BD simulations. Figure \ref{fig:binary_time}(c)--(f) compares the DDFT calculations to the BD simulations at intermediate (50\% of $t_{\rm f}$) and final drying times for the same conditions as in Figs.~\ref{fig:binary_time}(a)--(b). In qualitative agreement with the BD simulations, all three functionals predicted small-on-top stratification, with $\phi_{\rm S}$ being enhanced and $\phi_{\rm B}$ being decreased near the solvent interface. However, the volume fraction profiles determined using DDFT all had some quantitative disagreement with the BD simulations. The virial EOS gave more big particles (larger $\phi_{\rm B}$) at the top of the film than the simulations, while the BMCSL EOS and FMT gave fewer big particles (smaller $\phi_{\rm B}$); commensurately, the BMCSL EOS and FMT both overpredicted $\phi_{\rm B}$ in the bottom of the film. Unlike in the one-component suspensions, FMT and the BMCSL EOS both gave similar $\phi_{\rm S}$ and $\phi_{\rm B}$, and FMT was not substantially more accurate than the BMCSL EOS. The biggest difference between FMT and the BMCSL EOS was in $\phi_{\rm S}$ near the solvent interface, where FMT faithfully predicted the oscillations that the BMCSL EOS could not capture.

We attempted to quantify the accuracy of the different functionals by computing the root mean squared error in both $\phi_{\rm S}$ and $\phi_{\rm B}$ using a similar procedure as for the one-component suspensions (Figs.~S12 and S17). However, this analysis did not reveal any helpful trends because the various models all deviated from the BD simulations to a similar extent but in different ways. Accordingly, we have chosen not to discuss these measurements here. Instead, we characterized the extent of stratification in the two-component suspensions, which quantifies how much the small and big particles separate during drying. Several methods have been proposed to characterize this process, and we have used two different approaches here.

The first approach, which is common in the literature \cite{howard:lng:2017,tang:lng:2019,tang:jcp:2019,he:lng:2021}, characterizes the extent of stratification from the difference in the average displacement of the centers of mass of the small and big particles. The center of mass $\langle z_i \rangle$ of type $i$ can be computed from the density profile:
\begin{equation}
\langle z_i(t)\rangle = \frac{\int_{0}^{H(t)} \d{z} z \rho_i(z,t)}{\int_{0}^{H(t)} \d{z} \rho_i(z,t)}.
\label{eq:zcom}
\end{equation}
The displacement of the center of mass from its initial value is $\Delta\langle z_i\rangle = \langle z_i(t) \rangle - \langle z_i(0) \rangle$. By computing the difference in the displacements $\Delta\Delta z = \Delta \langle z_{\rm S} \rangle-\Delta \langle z_{\rm B} \rangle$, the extent of stratification can be characterized: a positive $\Delta\Delta z$ indicates small-on-top stratification because the big particles have displaced more (in the downward, negative direction) than the small particles.

We computed $\Delta\Delta z$ during drying for both our BD simulations and DDFT calculations. The top row of Figure \ref{fig:extent_of_strat} shows the results for various initial compositions for the same big-particle diameter and drying conditions as Fig.~\ref{fig:binary_time}. (We note that the BMCSL EOS is not included for the 3:1 initial composition [Figs.~\ref{fig:extent_of_strat}(e)--(f)] because the calculations failed numerically, see Sec.~\ref{sec:conclude}.) Initially, $\Delta\Delta z$ typically increased from zero as the film dried, consistent with the formation of a small-on-top layer. At the 1:1 and 3:1 initial compositions, the virial EOS somewhat underpredicted $\Delta\Delta z$, while the BMCSL EOS and FMT both overpredicted it. For the 1:3 initial composition, the BMCSL EOS and FMT predicted a negative $\Delta\Delta z$ in the early stages of drying, but all three functionals overpredicted $\Delta\Delta z$ by the end of drying. For all initial compositions, the BMCSL EOS and FMT gave nearly identical values of $\Delta\Delta z$, which is consistent with the similarity of $\phi_{\rm S}$ and $\phi_{\rm B}$.

We found, however, that there were a few drawbacks to using $\Delta\Delta z$ to characterize extent of stratification. First, although $\Delta\Delta z$ for the virial EOS was in reasonable agreement with the BD simulations, the profiles of $\phi_{\rm S}$ and $\phi_{\rm B}$ visually showed weaker stratification in the surface layer. Second, for all three initial compositions, $\Delta\Delta z$ was weakly nonmonotonic at late times. This behavior has also been observed by others \cite{he:lng:2021} but is somewhat inconsistent with the BD simulation trajectories or the volume fraction profiles, where the small-particle layer on the top of the film did not appear to shrink at the late stages of drying. Finally, $\Delta\Delta z$ had comparable values in the BD simulations for all initial compositions, although the small-on-top stratification was visually more pronounced for the 3:1 initial composition than the 1:3 initial composition. Together, this suggests that $\Delta\Delta z$ may not fully describe the stratification process. One potential reason for this is that the displacement is averaged over the entire film rather than near the surface where the effects of stratification are most dramatic.

Accordingly, we proposed a second approach to measure the extent of stratification that is based on the average surface composition by volume. The average fraction $\nu_i$ of type $i$ by volume in a surface layer of thickness $w$ was computed as
\begin{equation}
\nu_i(t) = \frac{1}{w} \int_{H(t)-w}^{H(t)} \d{z} \frac{\phi_i(z,t)}{\phi_{\rm S}(z,t)+\phi_{\rm B}(z,t)}.
\label{eq:surfvol}
\end{equation}
This quantity is similar to what can be measured in experiments using atomic force microscopy to probe the film surface \cite{Routh:2013}; however, in our simulations, we can in principle scan the composition in arbitrary volumes at arbitrary depth. We chose $w = 2\,d_{\rm B}$ to focus on the surface of the film.

We computed the difference in surface composition $\Delta \nu = \nu_{\rm S}-\nu_{\rm B}$, which ranges from $-1$ for 100\% big particles to $1$ for 100\% small particles (bottom row of Fig.~\ref{fig:extent_of_strat}). This approach was able to capture some of the behaviors that were missed using $\Delta\Delta z$. For example, $\Delta \nu$ increased monotonically for all methods and initial compositions. The small-particle surface layer formed slowest and had the weakest stratification for the 1:3 initial composition [Fig.~\ref{fig:extent_of_strat}(b)] and the strongest stratification for the 3:1 initial composition [Fig.~\ref{fig:extent_of_strat}(f)]. We found that $\Delta \nu$ was smaller for the virial EOS and larger for the BMCSL EOS and FMT than in the BD simulations. There was also a small difference in $\Delta\nu$ for the BMCSL EOS and FMT, which was not significant for $\Delta\Delta z$. We believe this difference is caused by differences in $\phi_{\rm S}$ near the solvent interface that are important when computing $\Delta \nu$ (localized to the surface) but are negligible when computing $\Delta \Delta z$ (averaged over the film).

We found that none of the hard-sphere functionals we tested, including FMT, was able to quantitatively predict the extent of stratification by either metric. However, the DDFT calculations were nonetheless satisfactory. The DDFT calculations predicted the occurrence of stratification, captured trends in the extent of stratification with respect to the initial composition and particle size, and were typically faster to perform than the BD simulations. The virial EOS was the least accurate of the LDA functionals we tested, consistent with the one-component suspensions, and it underpredicted the extent of stratification. The BMCSL EOS and FMT functionals both modestly overpredicted the extent of stratification in a comparable amount, and both might be similarly suitable for modeling stratification. Since FMT was highly accurate for the one-component suspensions, we speculate that some of the inaccuracy in DDFT for the two-component suspension may be due to the ``adiabatic'' assumption that particle correlations are given by an equivalent equilibrium system with the same density profile \cite{archer:jcp:2004,schmidt:jcp:2013}. This assumption might be relaxed using power-functional theory \cite{schmidt:jcp:2013,schmidt:rmp:2022} and is worthy of exploration in future. There may also be some numerical artifacts related to solving the DDFT equations that are more significant in two-component suspensions than in one-component suspensions, which we will comment on in Sec.~\ref{sec:conclude}.

\section{Conclusions}
\label{sec:conclude}
We have systematically tested the accuracy of various thermodynamic models for drying suspensions of hard-sphere colloidal particles. We compared the volume fraction profiles predicted using a continuum model based on classical DDFT to particle-based BD simulations for one-component and two-component films. Both models used the same free-draining approximation for the particle dynamics in the solvent, allowing us to assess the accuracy of the approximate free-energy functional that is an input to DDFT. We tested four free-energy functionals for hard-sphere particles: three LDA functionals based on the ideal-gas, virial, and BMCSL EOSs and one weighted-density approximation based on Rosenfeld's FMT.

For the one-component suspensions, we found that FMT was unambiguously the most accurate functional across a range of conditions, faithfully capturing not only the overall time evolution of the volume fraction profile but also oscillations in the volume fraction near the film interfaces. However, volume fraction profiles generated using the LDA with the BMCSL EOS were similar to those generated using FMT, except when oscillations developed at higher concentrations and in confinement, and might therefore be suitable for analysis outside these regimes. The LDA with the virial and ideal-gas EOSs were far less accurate except at low particle concentrations, where the underlying bulk EOSs are most accurate. The virial expansion has often been used to model hard-sphere suspensions out of equilibrium \cite{batchelor:jfm:1976,batchelor:jfm:1983,Brady:2011bh,zhou:prl:2017}, but we recommend caution in extrapolating these results to concentrated suspensions.

For the two-component suspensions of small and big particles, we found that none of the functionals tested clearly outperformed the others. Both FMT and the LDA with the BMCSL EOS overpredicted the extent of stratification by size compared to BD simulations, while the LDA with the virial EOS underpredicted the same. FMT was slightly more accurate than the BMCSL EOS because it better captured the small-particle structuring near the solvent interface; however, this structure was negligible in scale compared to the overall stratification by size in the film. Given the superior accuracy of FMT for the one-component suspensions, inaccuracy of FMT for the two-component suspensions could be caused by approximations of either the functionals for mixtures or the underlying DDFT itself; this point warrants further investigation.

In choosing a functional for DDFT, accuracy is not the sole factor that must be considered. More accurate models (such as FMT) tend to be more computationally demanding, and a balance must often be found between accuracy and performance. We conclude with a few computational considerations we noted using the various functionals. First, we emphasize that the DDFT calculations were uniformly more computationally efficient than the BD simulations. For the one-component suspensions, the DDFT calculations using FMT---the most computationally demanding functional---were between 3 and 13 times faster than the equivalent BD simulation using half the computational resources. This speedup is more impressive when we consider that multiple (5) BD simulations were needed to obtain statistically acceptable average profiles, while DDFT inherently produces the averaged result. The performance difference between DDFT using FMT and BD was reduced for the two-component suspensions, such that DDFT and BD had comparable total computational cost after factoring in both run time and resources; however, only one DDFT calculation was needed, instead of multiple BD simulations, to obtain a faithful average. Second, the DDFT calculations using the LDA functionals were significantly faster than using FMT; for the one-component suspensions, the LDA functions were typically 10 times faster than FMT (roughly 100 times faster than BD). This speedup is due to the reduced complexity of evaluating the LDA functionals, which do not require convolution of the density profiles, making the LDA functionals appealing for situations where quantitatively capturing fine details such as interfacial structuring is not critical. Based on Figs.~\ref{fig:binary_time} and \ref{fig:extent_of_strat}, this approximation may be especially useful for two-component suspensions where interfacial effects may be negligible compared to film-scale composition gradients.

Last, however, we found that the DDFT calculations using the LDA functionals failed under certain conditions. For example, the DDFT calculation for the one-component suspension with $\phi_{0,{\rm S}} = 0.05$ and $H_0 = 100 d_{\rm S}$ failed when the overall suspension concentration approached $\phi_{\rm f,{\rm S}} \approx  0.5$. Similar challenges were also encountered in the two-component suspensions, where the DDFT calculations using the BMCSL EOS failed to complete for $d_{\rm B} = 3\,d_{\rm S}$ with 3:1 initial composition and so were omitted from Figs.~\ref{fig:binary_time}(e)--(f). Perhaps most dramatically, none of the DDFT calculations using the LDA functionals completed for $d_{\rm B} = 6\,d_{\rm S}$, while calculations using FMT were successful for the 1:3 and 1:1 initial compositions (Figs.~S13--S14) but failed near the end for the 3:1 initial composition (Fig.~S15). Failure of these calculations typically appeared as a numerical instability when evolving Eq.~\eqref{eq:ddft}. Hence, in addition to its increased accuracy, there may be some numerical advantages to using a weighted-density approximation like FMT that helps outweigh some of its increased computational costs. Further investigation of the numerical scheme used to integrate Eq.~\eqref{eq:ddft}, particularly for the LDA functionals, would be helpful.

This study has explored the accuracy of different free-energy functionals for modeling inhomogeneous suspensions of hard-sphere particles using DDFT. This information provides guidance for selecting thermodynamic models for not only the drying process we studied but also related confined nonequilibrium processes such as solvent freezing \cite{xu:2019} or sedimentation \cite{royall:prl:2007}. We have focused on modeling hard-sphere particles, but the functionals we studied can be incorporated into thermodynamic perturbation theories \cite{Weeks:1971} to model particles with other types of interactions. We have also made a consistent free-draining approximation of the particle dynamics in both the DDFT calculations and BD simulations, but there is evidence that solvent-mediated hydrodynamic interactions between particles play an important role in setting the structure of drying suspensions \cite{sear:pre:2017,statt:jcp:2018,howard:jcp:2020,Park:2022}. Faithfully modeling these dynamic effects in DDFT \cite{rex:prl:2008,rex:epje:2009} ultimately still requires an accurate underlying free-energy functional, so this work constitutes an important step in building improved DDFT models for drying suspensions.

\begin{acknowledgements}
This work was completed with resources provided by the Auburn University Easley Cluster.
\end{acknowledgements}

\section*{Data Availability}
The data that support the findings of this study are available from the authors upon reasonable request.

\bibliography{references.bib}

\end{document}


\title{Supplementary material for ``Dynamic density functional theory for drying colloidal suspensions: Comparison of hard-sphere free-energy functionals''}

\author{Mayukh Kundu}
\affiliation{Department of Chemical Engineering, Auburn University, Auburn, AL 36849, USA}

\author{Michael P. Howard}
\email{mphoward@auburn.edu}
\affiliation{Department of Chemical Engineering, Auburn University, Auburn, AL 36849, USA}

\maketitle

\section{One-component suspension}
\subsection{Initial film height $H_0 = 100\,d_{\rm S}$}
\begin{figure}[!h]
  \centering
  \includegraphics{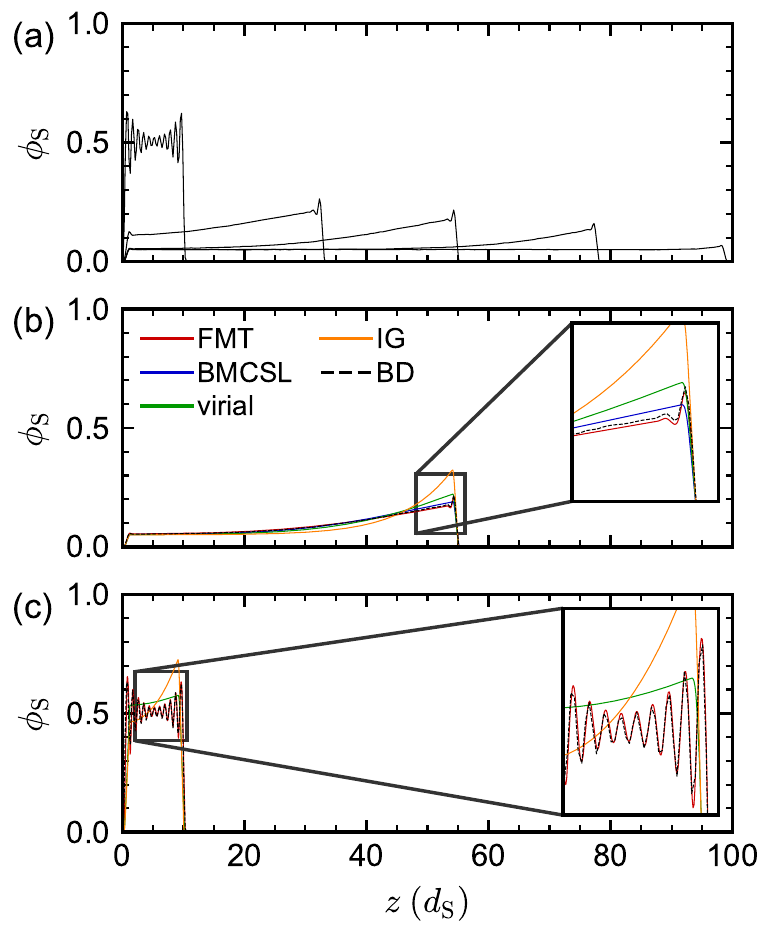}
  \caption{Same as Fig.~2 with $\phi_{0,{\rm S}} = 0.05$.}
  \label{fig:onecomp:h100phi5}
\end{figure}

\begin{figure}[!h]
  \centering
  \includegraphics{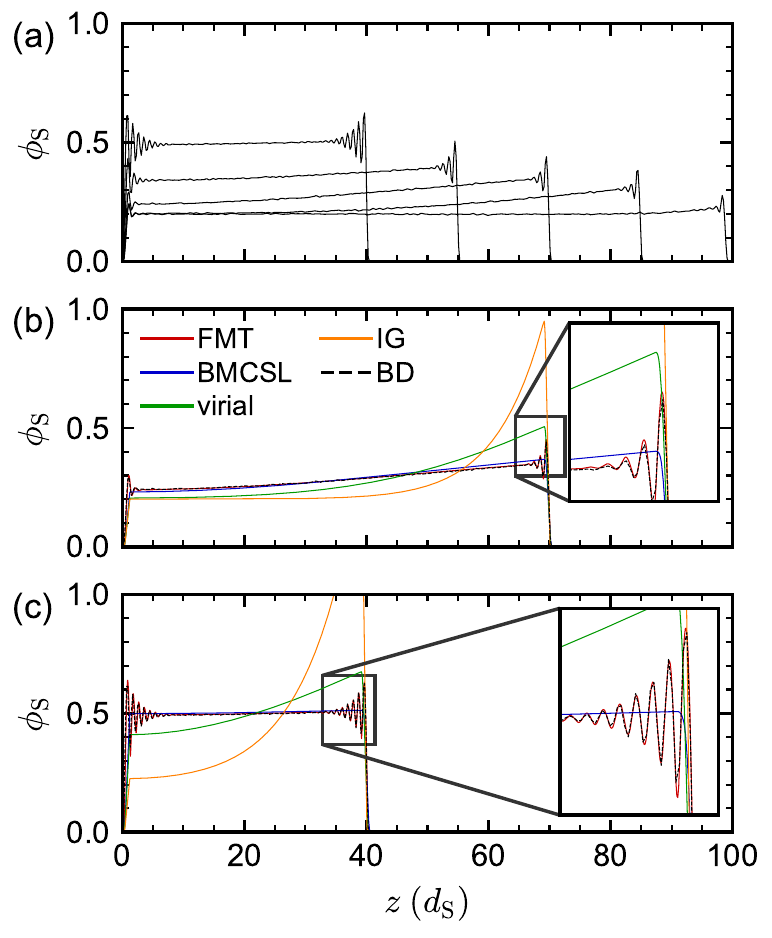}
  \caption{Same as Fig.~2 with $\phi_{0,{\rm S}} = 0.20$.}
  \label{fig:onecomp:h100phi20}
\end{figure}

\begin{figure}[!h]
  \centering
  \includegraphics{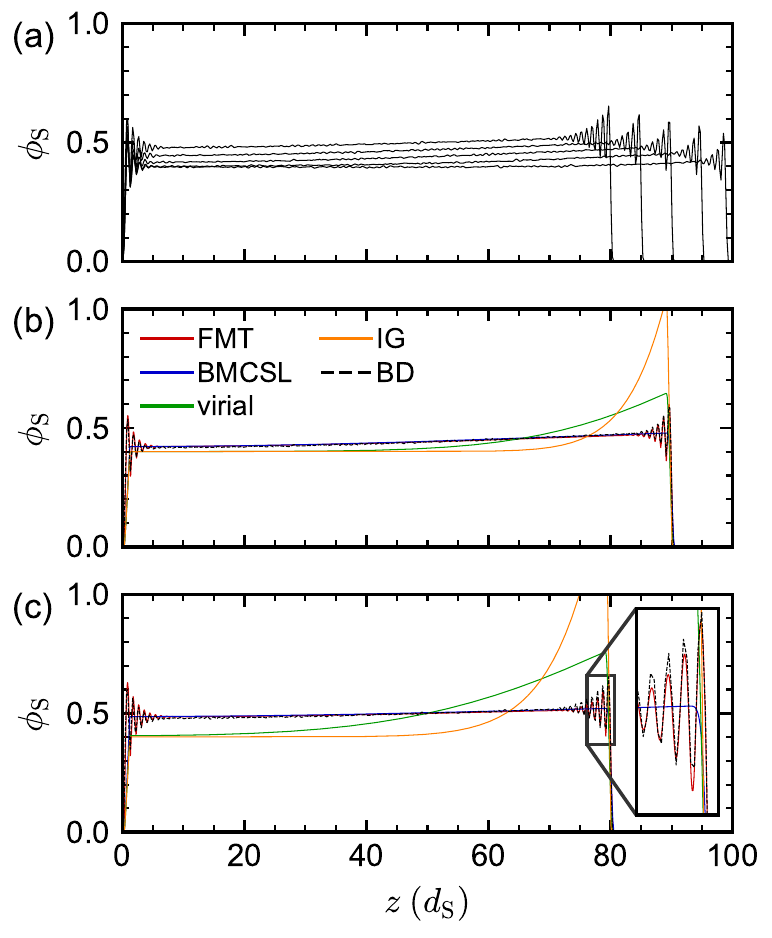}
  \caption{Same as Fig.~2 with $\phi_{0,{\rm S}} = 0.40$.}
  \label{fig:onecomp:h100phi40}
\end{figure}

\clearpage
\subsection{Initial film height $H_0 = 50\,d_{\rm S}$}
\begin{figure}[!h]
    \includegraphics{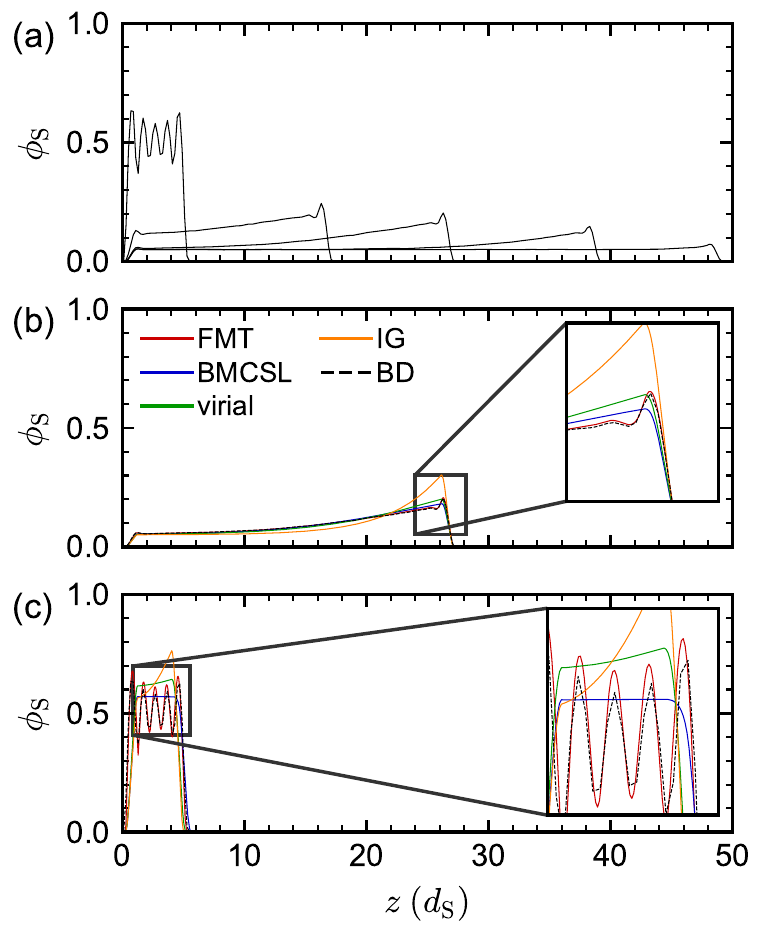}
    \caption{Small-particle volume fraction profile $\phi_{\rm S}$ in drying one-component suspension with $\phi_{0,{\rm S}} = 0.05$, $H_0 = 50\,d_{\rm S}$, and $v = 0.2\,d_{\rm S}/\tau$. (a) BD simulation at times corresponding to 0\%, 25\%, 50\%, 75\% and 100\% of the total drying time $t_{\rm f}$. (b)--(c) Comparison of DDFT calculations with different free-energy functionals to BD simulation at (b) 50\% of $t_{\rm f}$ and (c) 100\% of $t_{\rm f}$.}
    \label{fig:onecomp:h50phi5}
\end{figure}

\begin{figure}[!h]
  \centering
  \includegraphics{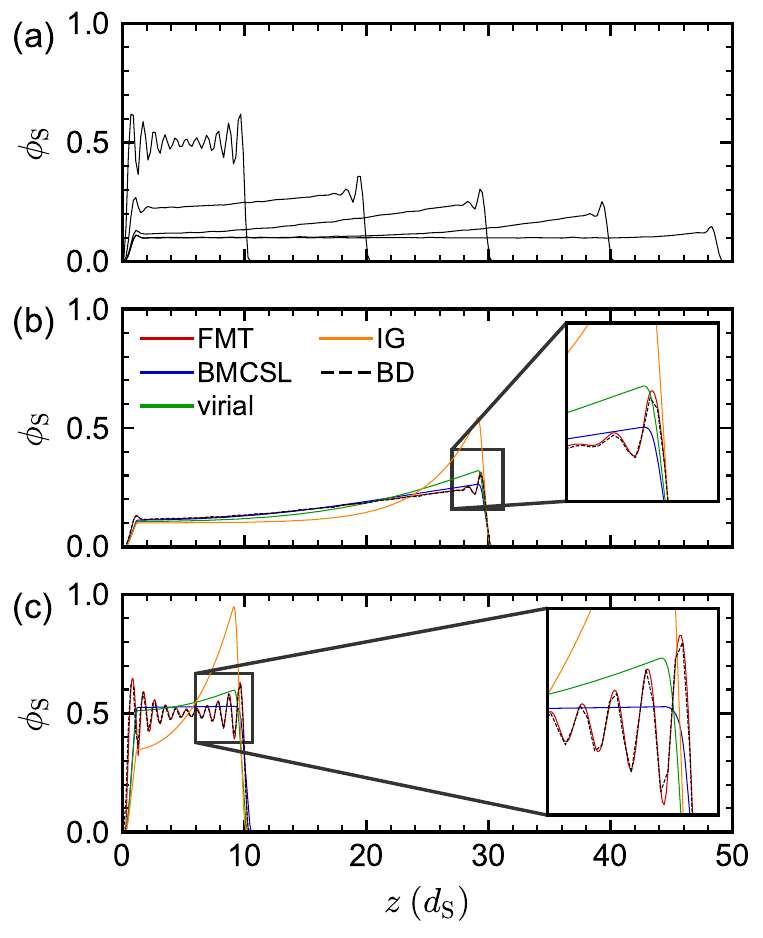}
  \caption{Same as Fig.~\ref{fig:onecomp:h50phi5} with  $\phi_{0,{\rm S}} = 0.10$.}
  \label{fig:onecomp:h50phi10}
\end{figure}

\begin{figure}[!h]
  \centering
  \includegraphics{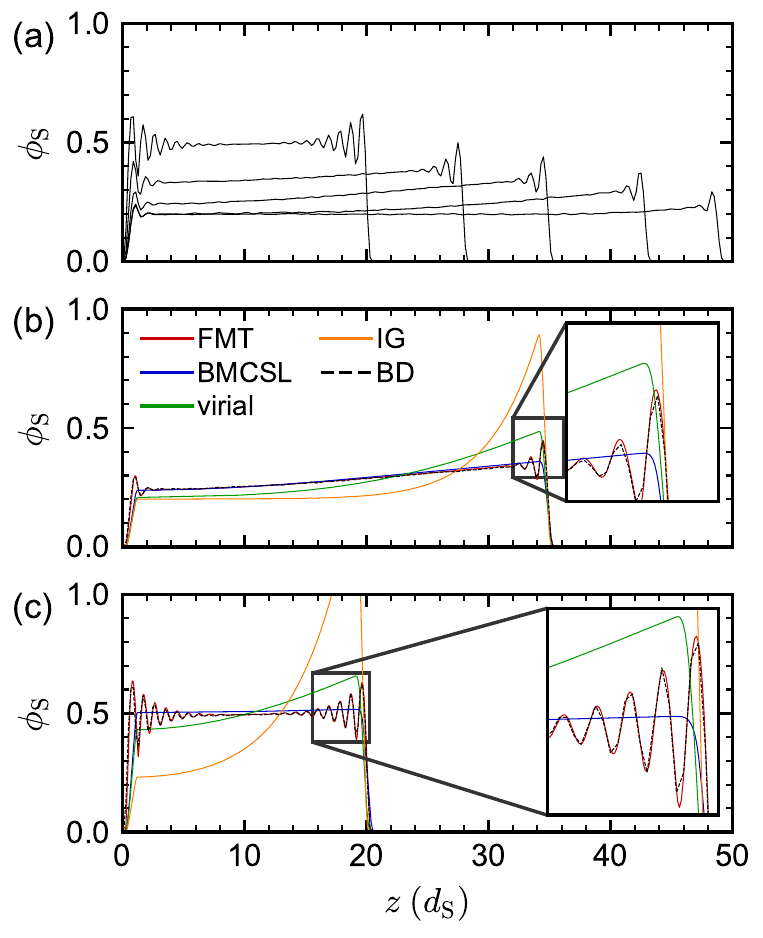}
  \caption{Same as Fig.~\ref{fig:onecomp:h50phi5} with  $\phi_{0,{\rm S}} = 0.20$.}
  \label{fig:onecomp:h50phi20}
\end{figure}

\begin{figure}[!h]
  \centering
  \includegraphics{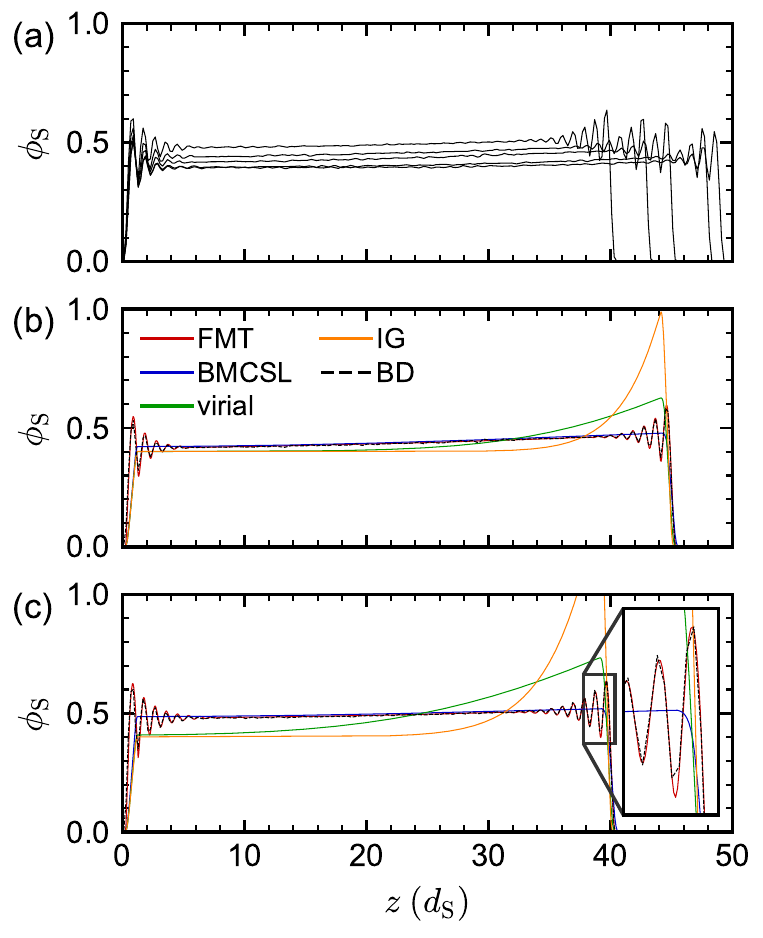}
  \caption{Same as Fig.~\ref{fig:onecomp:h50phi5} with  $\phi_{0,{\rm S}} = 0.40$.}
  \label{fig:onecomp:h50phi40}
\end{figure}

\begin{figure}[!h]
  \centering
  \includegraphics{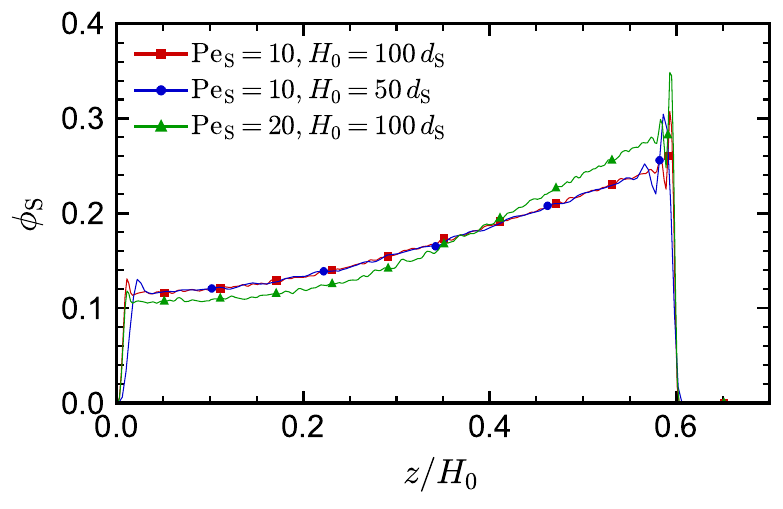}
  \caption{Small-particle volume fraction profiles from BD simulations for $\phi_{0,{\rm S}} = 0.10$ when films are dried to 50\% of $t_{\rm f}$, with position scaled by initial film height $H_0$. The profiles for the two drying conditions having the same film P\'{e}clet number ($H_0 = 100\,d_{\rm S}$ and $v = 0.1\,d_{\rm S}/\tau$, $H_0 = 50\,d_{\rm S}$ and $v=0.2\,d_{\rm S}/\tau$) essentially collapse, while the two drying conditions having the same drying rate ($H_0 = 100\,d_{\rm S}$ and $v = 0.2\,d_{\rm S}/\tau$, $H_0 = 50\,d_{\rm S}$ and $v=0.2\,d_{\rm S}/\tau$) do not.}
\end{figure}

\begin{figure}[!h]
    \centering
    \includegraphics{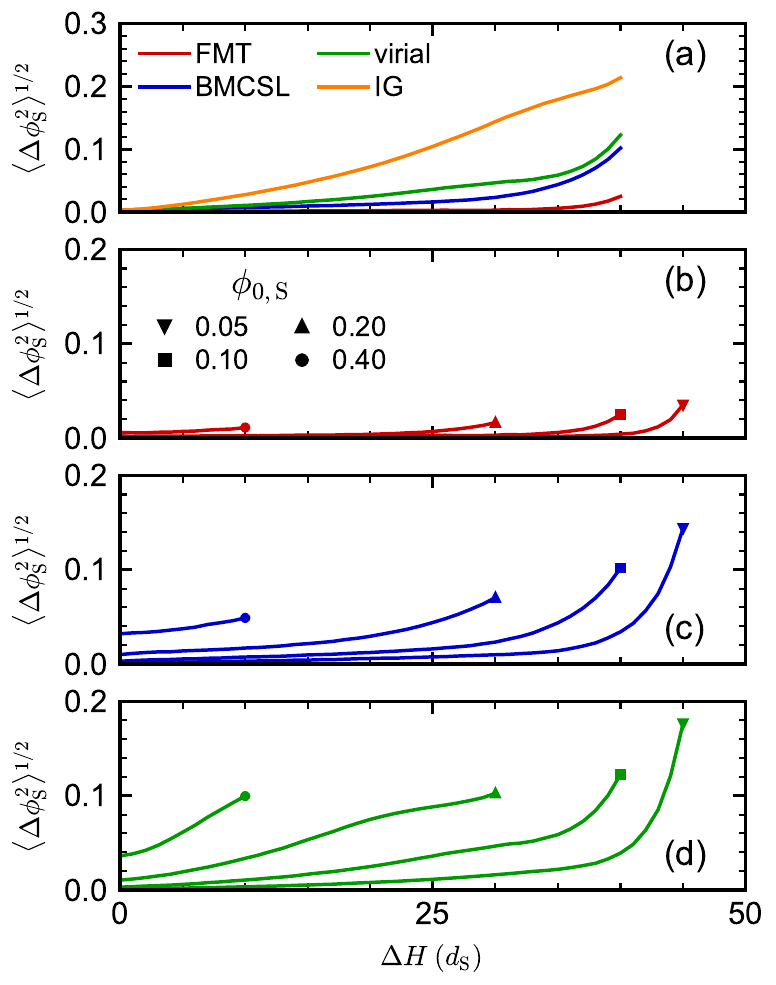}
    \caption{Root mean squared error $\langle \Delta \phi_{\rm S} \rangle^{1/2}$ in DDFT calculations of small-particle volume fraction profiles compared to BD simulations as a function of solvent interface displacement $\Delta H = H_0-H(t)$  when $H_0 = 50\,d_{\rm S}$ and $v = 0.2\,d_{\rm S}/\tau$. (a) Comparison of error for the various free-energy functionals for $\phi_{0,{\rm S}} = 0.10$, as in Fig.~\ref{fig:onecomp:h50phi10}. (b)--(d) Error as initial volume fraction $\phi_{0,{\rm S}}$ is varied for (b) FMT and LDA using the (c) BMCSL EOS and (d) virial EOS.}
    \label{fig:single_error}
\end{figure}

\clearpage
\section{Two-component suspension}
\subsection{Big particle diameter, $d_{\rm B} = 3 d_{\rm S}$}
\begin{figure*}[!h]
    \centering
    \includegraphics{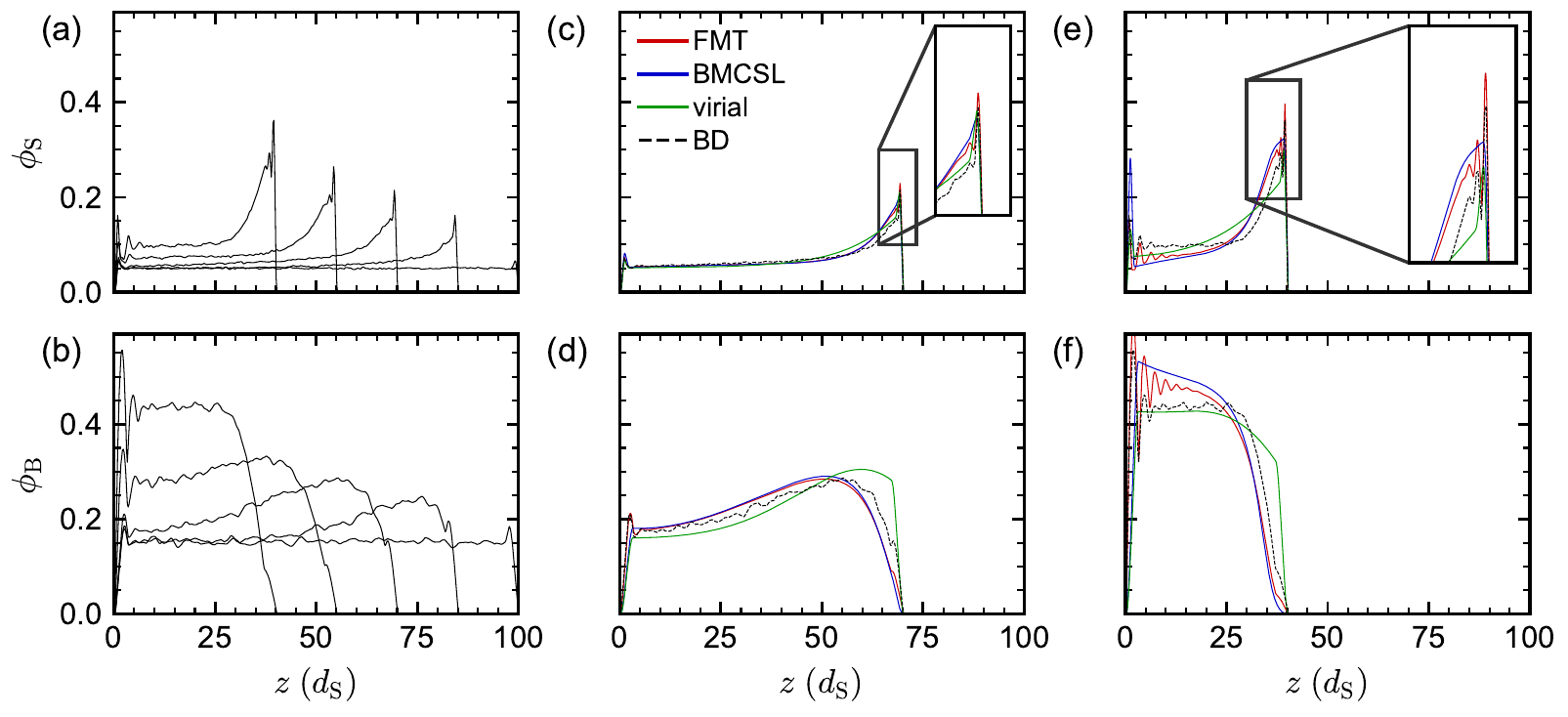}
    \caption{Same as Fig.~5 for 1:3 initial composition ($\phi_{0,{\rm S}} = 0.05$, $\phi_{0,{\rm B}} = 0.15$).}
    \label{fig:twocomp:dB3:h100phiS05}
\end{figure*}

\begin{figure*}[!h]
    \centering
    \includegraphics{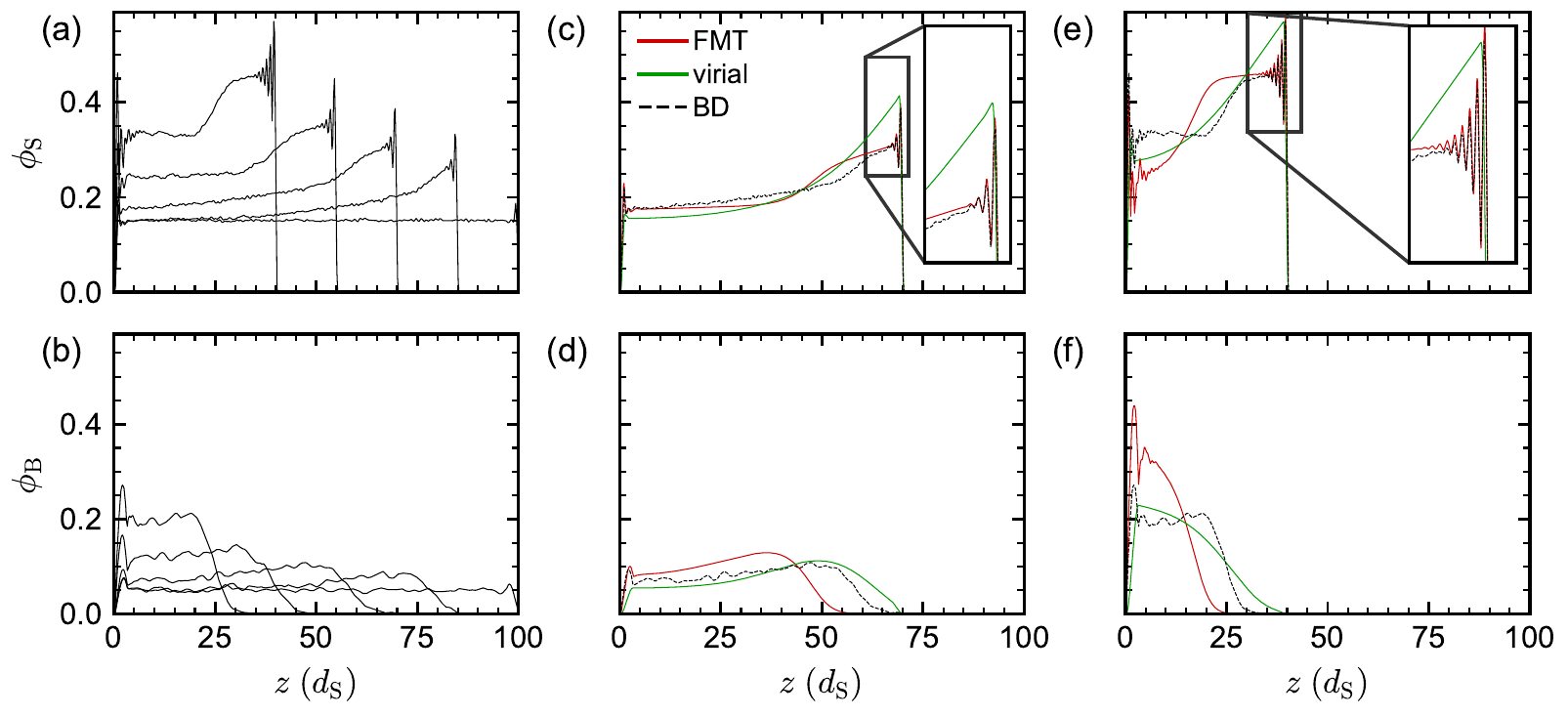}
    \caption{Same as Fig.~5 for 3:1 initial composition ($\phi_{0,{\rm S}} = 0.15$, $\phi_{0,{\rm B}} = 0.05$). Note that the BMCSL result is not included because of numerical failure at roughly 75\% of the total drying time.}
    \label{fig:twocomp:dB3:h100phiS15}
\end{figure*}

\begin{figure}[!h]
    \centering
    \includegraphics{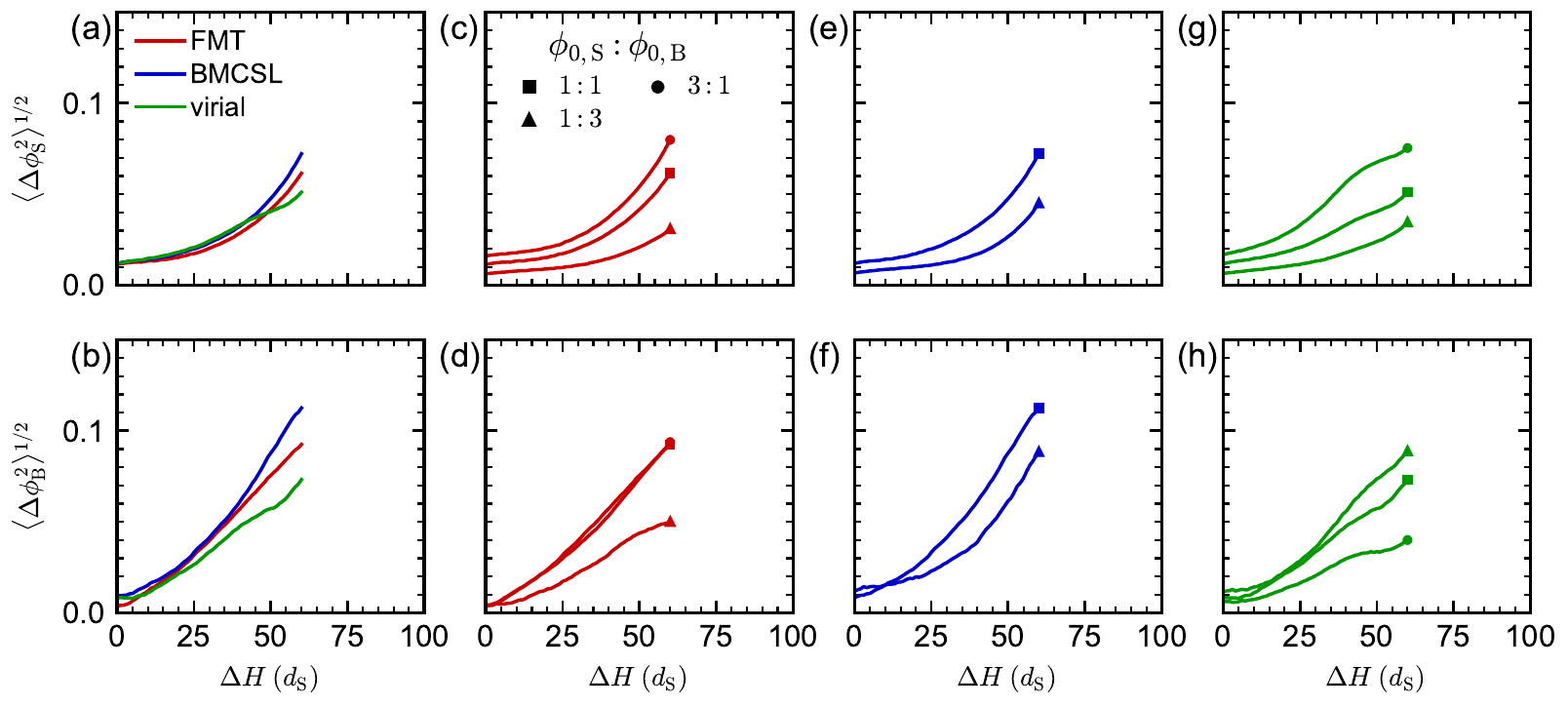}
    \caption{Root mean squared error $\langle \Delta \phi_{\rm S} \rangle^{1/2}$ and $\langle \Delta \phi_{\rm B} \rangle^{1/2}$ in DDFT calculations of small- and big-particle volume fraction profiles, respectively, compared to BD simulations as a function of solvent interface displacement $\Delta H = H_0-H(t)$ when $d_{\rm B} = 3\,d_{\rm S}$, $H_0 = 100\,d_{\rm S}$, and $v=0.1\,d_{\rm S}/\tau$. (a)--(b) Comparison of error for the various free-energy functionals for the same conditions as Fig.~5. (c)--(h) Error as initial composition is varied for (c)--(d) FMT and LDA using the (e)--(f) BMCSL EOS and (g)--(h) virial EOS.}
    \label{fig:twocomp:dB3:error}
\end{figure}

\clearpage
\subsection{Big particle diameter, $d_{\rm B} = 6 d_{\rm S}$}
\begin{figure*}[!h]
    \centering
    \includegraphics{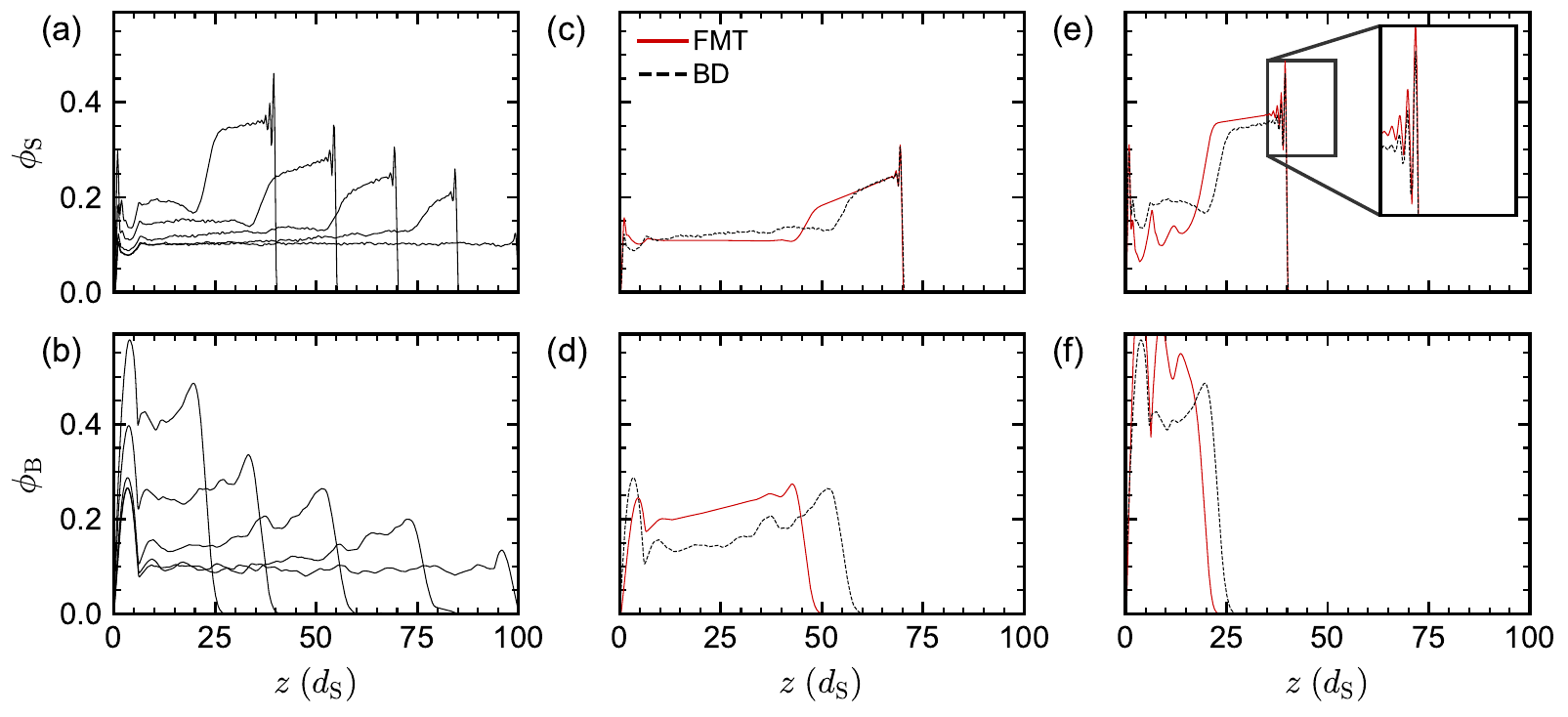}
    \caption{Same as Fig.~5 with $d_{\rm B} = 6\,d_{\rm S}$. Note that the DDFT calculation is shown only for the FMT functional due to numerical failure of all calculations with the LDA functionals.}
    \label{fig:twocomp:dB6:h100phiS1}
\end{figure*}

\begin{figure*}[!h]
    \centering
    \includegraphics{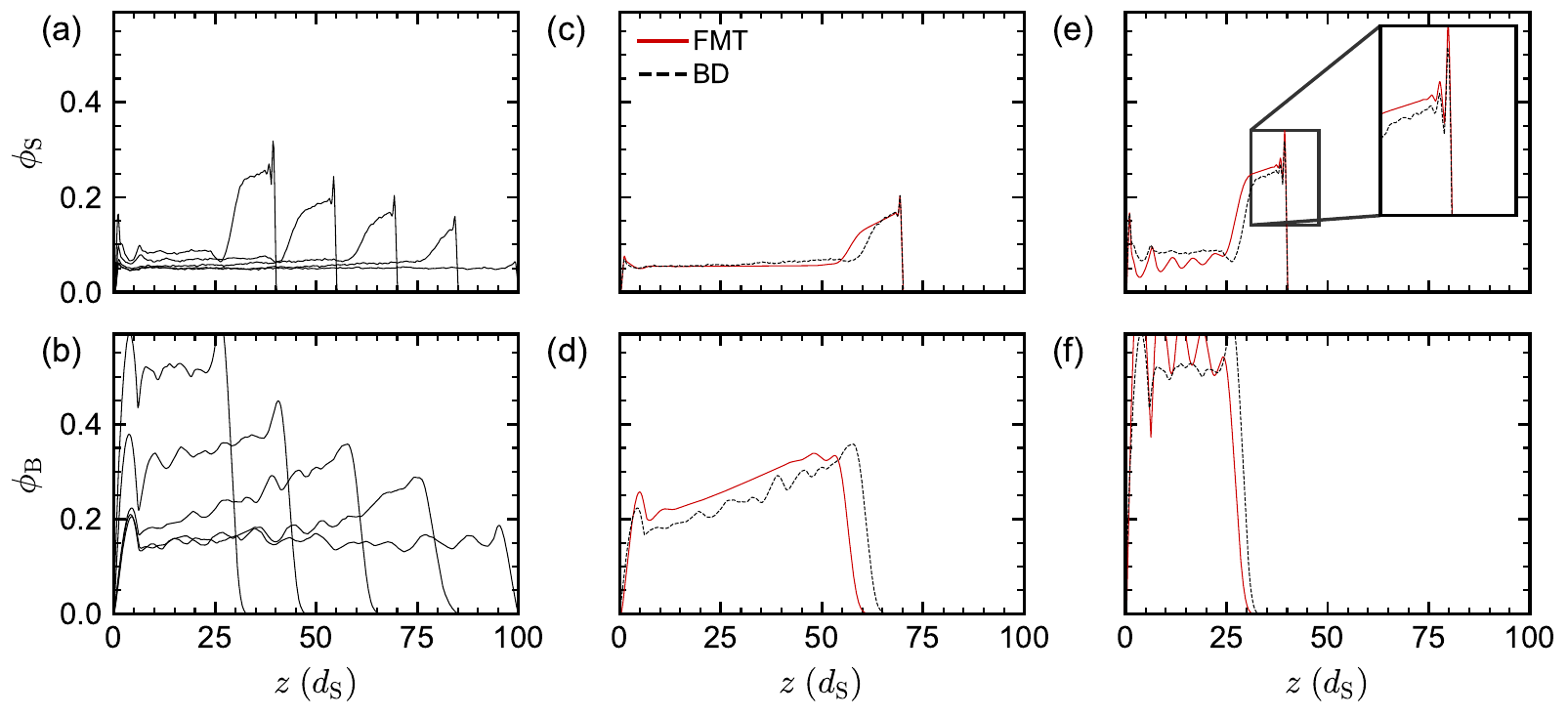}
    \caption{Same as Fig.~\ref{fig:twocomp:dB6:h100phiS1} for 1:3 initial composition ($\phi_{0,{\rm S}} = 0.05$, $\phi_{0,{\rm B}} = 0.15$).}
    \label{fig:twocomp:dB6:h100phiS05}
\end{figure*}

\begin{figure*}[!h]
    \centering
    \includegraphics{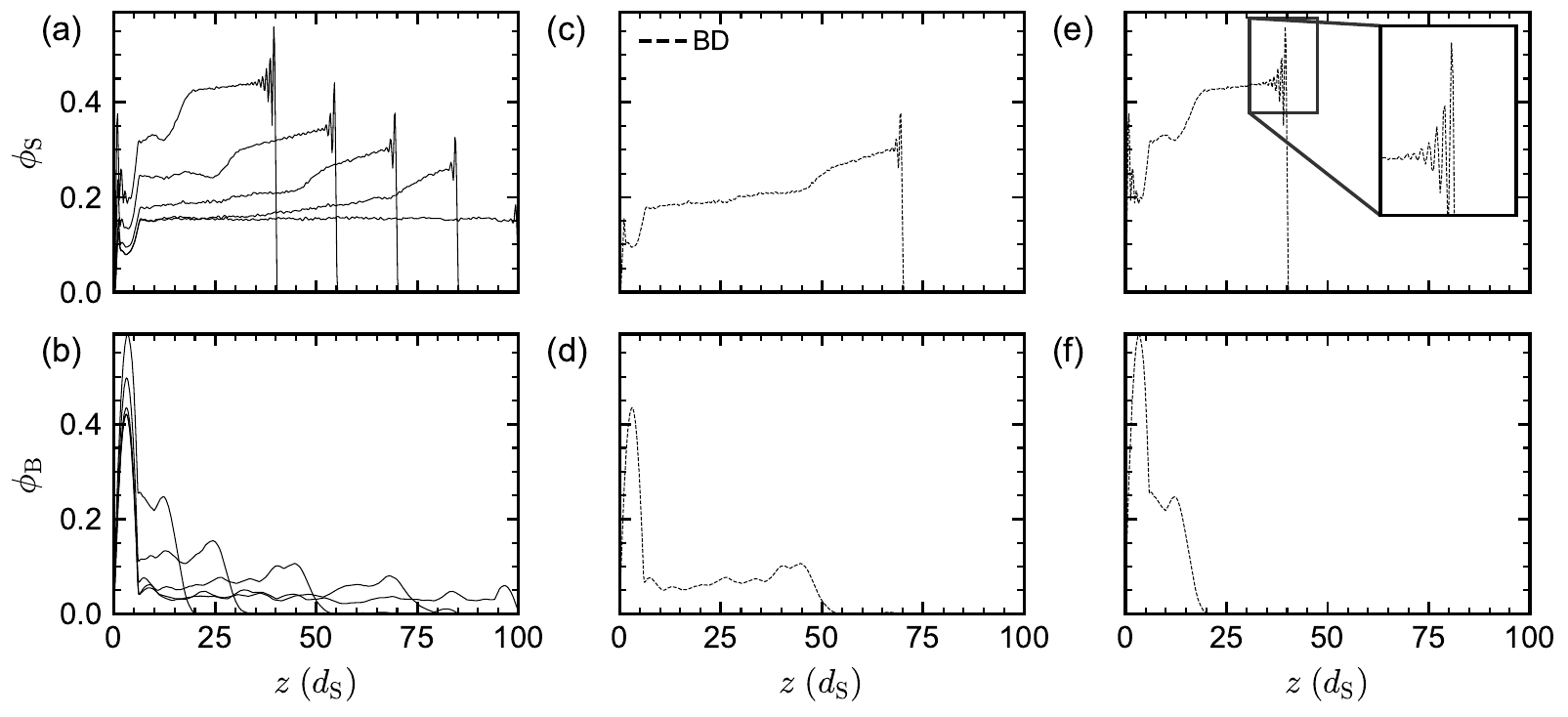}
    \caption{Same as Fig.~\ref{fig:twocomp:dB6:h100phiS1} for 3:1 initial composition ($\phi_{0,{\rm S}} = 0.15$, $\phi_{0,{\rm B}} = 0.05$), but only showing BD results due to numerical failure of the DDFT calculations with the FMT functional after nearly 90\% of the total drying time.}
    \label{fig:twocomp:dB6:h100phiS15}
\end{figure*}

\begin{figure}[!h]
    \includegraphics{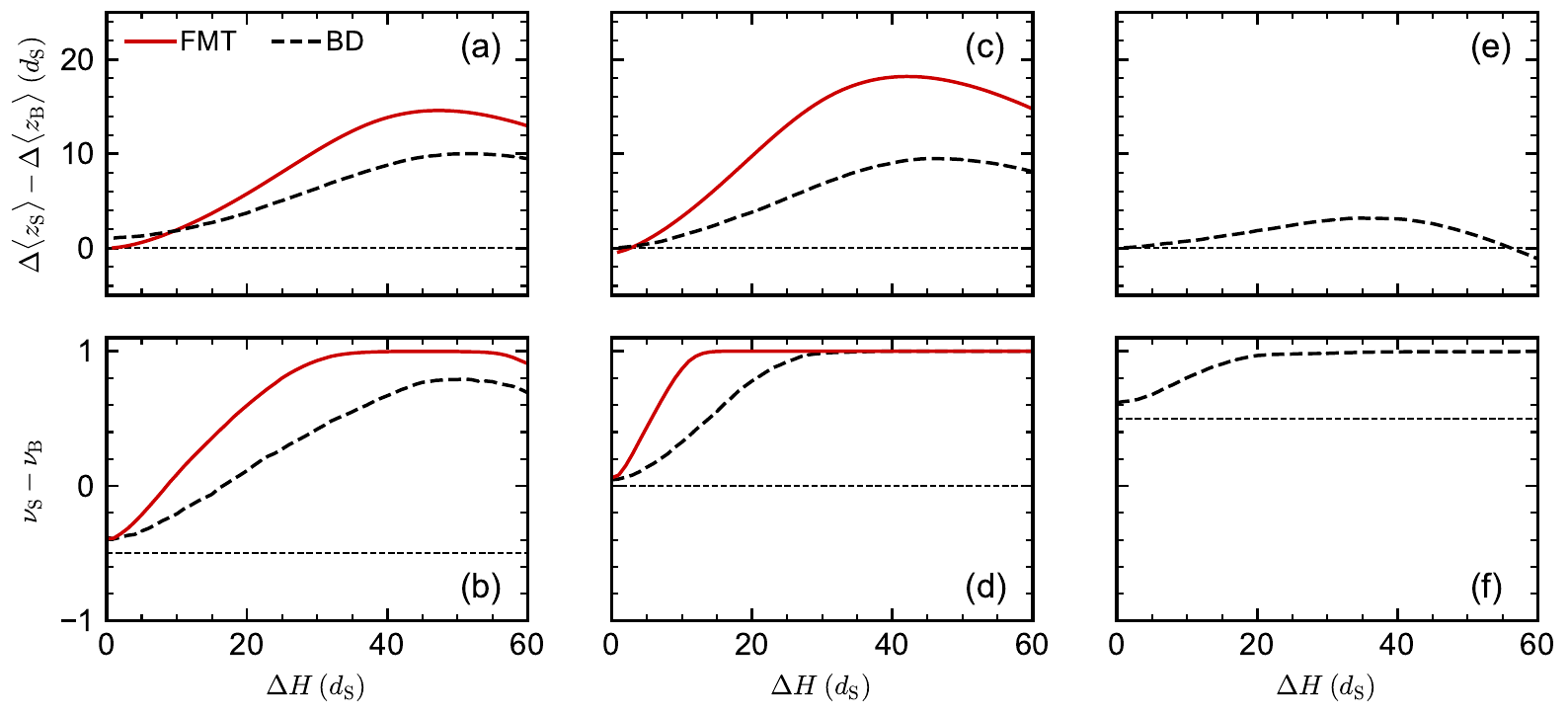}
    \caption{Same as Fig.~7 with $d_{\rm B} = 6\,d_{\rm S}$.}
    \label{fig:extent_of_strat_dB_6}
\end{figure}

\begin{figure}[!h]
    \centering
    \includegraphics{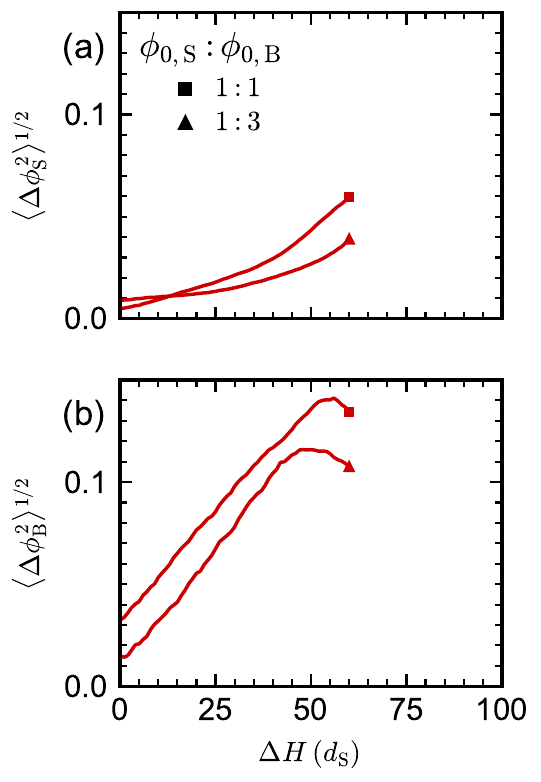}
    \caption{Same as Fig.~\ref{fig:twocomp:dB3:error} with $d_{\rm B} = 6\,d_{\rm S}$.}
    \label{fig:twocomp:dB6:error}
\end{figure}